\documentclass[twocolumn,trackchanges,floatfix,twocolappendix,longbib]{aastex701}
\usepackage{textcomp}
\usepackage{amsmath}
\usepackage{longtable}
\usepackage{multirow}
\usepackage{booktabs}
\usepackage[english]{babel}
\usepackage{url}
\usepackage{natbib}
\usepackage{gensymb}
\usepackage{subfigure}
\DeclareUnicodeCharacter{2212}{-}
\usepackage{amsmath,bm}
\usepackage{amssymb}
\usepackage{multirow}
\usepackage{threeparttable}
\usepackage{times}
\usepackage{enumitem}
\usepackage{lipsum}[1-2]
\usepackage{array}

\newcommand{\Gaia}{\textit{Gaia}}

\shorttitle{Candidate CEMP Stars of the Galaxy}
\shortauthors{Hong et al.}
\begin{document}

\title{Carbon-Enhanced Metal-Poor Star Candidates in the Milky Way from J-PLUS and S-PLUS}

\author[0000-0002-2453-0853]{Jihye Hong}
\affiliation{Department of Physics and Astronomy, University of Notre Dame, Notre Dame, IN 46556, USA}
\affiliation{Joint Institute for Nuclear Astrophysics -- Center for the Evolution of the Elements (JINA-CEE), USA}
\email{jhong5@nd.edu}

\author[0000-0003-4573-6233]{Timothy C. Beers}
\affiliation{Department of Physics and Astronomy, University of Notre Dame, Notre Dame, IN 46556, USA}
\affiliation{Joint Institute for Nuclear Astrophysics -- Center for the Evolution of the Elements (JINA-CEE), USA}
\email{Timothy.C.Beers.5@nd.edu}

\author[0000-0003-3250-2876]{Yang Huang}
\affiliation{School of Astronomy and Space Science, University of Chinese Academy of Sciences, Beijing 100049, People's Republic of China}
\affiliation{National Astronomical Observatories, Chinese Academy of Sciences, Beijing 100101, People's Republic of China}
\email{huangyang@ucas.ac.cn}

\author[0009-0006-7257-913X]{Jonathan Cabrera Garcia}
\affiliation{Department of Physics and Astronomy, University of Notre Dame, Notre Dame, IN 46556, USA}
\affiliation{Joint Institute for Nuclear Astrophysics -- Center for the Evolution of the Elements (JINA-CEE), USA}
\email{jcabrer2@nd.edu}

\author[0000-0001-5297-4518]{Young Sun Lee}
\affiliation{Department of Astronomy and Space Science, Chungnam National University, Daejeon 34134, Republic of Korea}
\email{youngsun@cnu.ac.kr}

\author[0000-0003-4479-1265]{Vinicius M. Placco}
\affiliation{NSF NOIRLab, Tucson, AZ 85719, USA}
\email{vinicius.placco@noirlab.edu}

\author[0000-0001-6196-5162]{Evan N. Kirby}
\affiliation{Department of Physics and Astronomy, University of Notre Dame, Notre Dame, IN 46556, USA}
\email{ekirby@nd.edu}

%\date{\today}
\submitjournal{ApJ}
%\accepted{Apr18}

\begin{abstract}

\noindent Recent large-scale multi-band photometric surveys now enable elemental-abundance estimates for millions of stars with accuracies approaching those of low- to medium-resolution spectroscopy. Using [Fe/H] and [C/Fe] estimates derived from the Javalambre Photometric Local Universe Survey (J-PLUS) DR3 and the Southern Photometric Local Universe Survey (S-PLUS) DR4, which together cover $\sim$6,200 deg$^2$ of the sky, we identify large numbers of carbon-enhanced metal-poor (CEMP) stars in the Milky Way. After applying data-quality cuts and evolutionary corrections to the carbon-abundance estimates, we construct a combined J/S-PLUS sample of $\sim$6.40 million stars and identify $\sim$104,900 CEMP candidates, roughly twice the number of CEMP candidates identified from Gaia XP spectra by Lucey et al. We photometrically confirm that the absolute carbon abundance $A$(C) separates CEMP stars into two primary groups, CEMP-no and CEMP-$s$ stars, consistent with previous spectroscopic studies. We also recover CEMP morphological Groups I-III in the Yoon-Beers diagram, as well as the recently proposed Group IV, and show that it is statistically distinct even in photometric data. A cumulative frequency analysis confirms that the CEMP fraction increases toward lower metallicity and that CEMP-no stars dominate in the most metal-poor regime. By comparing frequencies with and without Group IV stars, we assess their relation to CEMP-no and CEMP-$s$ stars, and examine CEMP distributions across different Galactic components. The resulting catalog provides a substantial sample for future spectroscopic follow-up, in particular to constrain the likely origin(s) of the Group IV stars.

\end{abstract}


\section{Introduction}\label{sec:introduction}

Studies of metal-poor (MP; [Fe/H] $\leq –1$), very metal-poor (VMP; [Fe/H] $\leq –2$), and extremely metal-poor (EMP; [Fe/H] $\leq –3$) stars in the Milky Way (MW) and other galaxies provide important information on the ancient environments in which these stars formed. As stellar metallicity decreases, metal-deficient stars have been observed to exhibit higher carbon-to-iron abundance ratios, a trend that has been reported in numerous studies \citep[e.g.,][]{Luck1991, Beers2005, Cohen2005, Rossi2005, Frebel2006a, Lucatello2006, Aoki2007a, Carollo2012, Lee2013, Aoki2013, Yong2013b, Placco2014c, Beers2017, Yoon2018, Arentsen2022}. Among MP stars, those with [C/Fe] $> +1.0$ were initially designated as carbon-enhanced metal-poor (CEMP) stars by \citet{Beers2005}. \citet{Aoki2007a} later refined this classification according to stellar luminosity, adopting a [C/Fe] $> +0.7$ threshold for unevolved stars (log$(L/L_{\odot}) \leq 2.3$) and a luminosity-dependent threshold of [C/Fe] $> 3.0 – $log$(L/L_{\odot})$ for more luminous, evolved stars. \citet{Placco2014c} further developed a surface gravity–dependent correction procedure based on stellar-evolution models, which allows all stars to be classified as CEMP using a uniform [C/Fe] $> +0.7$ criterion.

CEMP stars can be further sub-classified according to their abundances of neutron-capture elements \citep{Beers2005, Frebel2018}. For example, the presence or absence of over-abundances in barium (Ba) and europium (Eu), which are predominantly produced via the slow ($s$-) and rapid ($r$-) neutron-capture processes, respectively, allows for a distinction between CEMP-$s$ stars ([Ba/Fe] $> +1.0$ and [Ba/Eu] $> +0.5$) and CEMP-no stars ([Ba/Fe] $< 0.0$). Additionally, CEMP-$r$/$s$ stars are characterized by $0 <$ [Eu/Ba] $\leq +0.5$, and CEMP-$r$ stars are defined as having [Eu/Fe] $> +0.7$ and [Ba/Eu] $< 0$.

These subclasses offer clues to their origins. For instance, the great majority of CEMP-$s$ stars likely formed in binary systems where mass transfer from a former asymptotic giant branch (AGB) companion -- now a white dwarf -- enriched the observed star with carbon and $s$-process elements such as Ba \citep{Suda2004, Herwig2005, Lucatello2005, Komiya2007, Campbell2008, Bisterzo2011, Placco2013, Starkenburg2014, Abate2015, Placco2015b, Hansen2016b, Yoon2016, Arentsen2019, Lee2019, Bonifacio2025}. In addition, particularly at very low metallicities, rotating massive stars have been proposed as viable contributors to carbon and the $s$-process elements \citep{Frischknecht2016, Choplin2018, Limongi2018}.

In contrast, CEMP-no stars, more frequently found to be single stars \citep{Hansen2016a}, are thought to be descendants of first-generation population III (Pop III) stars that produced C-rich but Fe-poor ejecta. One scenario invokes faint supernovae (SNe) from metal-free massive stars that underwent mixing and fallback, enriching the natal clouds of CEMP-no stars with C and other light elements such as N and O \citep{Umeda2003, Umeda2005, Iwamoto2005, Joggerst2009, Nomoto2013, Tominaga2014, Ishigaki2014b, Marassi2014, Salvadori2015, Komiya2020}.  Another explanation involves fast-rotating massive stars (spinstars), in which internal mixing and strong stellar winds enrich the surrounding interstellar medium with C and neutron-capture elements \citep{Meynet2006, Meynet2010, Chiappini2013, Maeder2015, Liu2021}.  

The origin of CEMP-$r$/$s$ stars remains unclear. One possibility is that they formed from gas clouds pre-enriched in $r$-process elements and were later enriched in $s$-process material by an AGB companion \citep{Jonsell2006, Abate2016, Bonifacio2025}.  The intermediate neutron-capture process ($i$-process), originally suggested by \citet{Cowan1977}, has also been proposed for the production of neutron-capture elements \citep{Hampel2019,choplin2024}.  It is likely to be associated with low-metallicity AGB stars \citep{choplin2024} and/or rapidly accreting white dwarfs \citep[][and references therein]{Denissenkov2019}.  Only a few tens of CEMP-$r/s$ stars have been recognized at present, but among those that are known, the binary fractions appear to be high, at least $\sim$ 50\% \citep{Yoon2016}. Collapsar disk winds \citep{Miller2020,Barnes2022} could generate substantial $r$-process yields, and have also been argued to be prodigious producers of carbon \citep{Siegel2019}, thus they also may be responsible for the origin of CEMP-$r$ stars. 

\citet{Rossi2005} originally pointed out that CEMP stars might not share a single nucleosynthetic origin, based on the apparent bimodal distribution of [C/Fe] and absolute carbon abundance $A$(C)\footnote{$A$(C) = $\log \epsilon(\mathrm{C}) = \log (N_\mathrm{C}/N_\mathrm{H}) + 12$, where $N$ denotes the number densities of atoms of carbon and hydrogen, respectively.} at low metallicity (see their Figure 12). Later,  \citet{spite2013} and \citet{Bonifacio2015} presented evidence that there existed high and low ``bands'' in plots of $A$(C) versus [Fe/H] for main-sequence turnoff stars and mildly evolved subgiants, primarily associated with CEMP-$s$ and CEMP-no stars, respectively.  

The full richness of the behavior of CEMP stars in the $A$(C)–[Fe/H] space was revealed in Figure 1 of \citet{Yoon2016} -- the so-called Yoon–Beers diagram -- which separated CEMP stars into three morphological groups. Yoon et al.\ incorporated newer, larger samples and applied carbon corrections from \citet{Placco2014c} that account for stellar evolutionary stages. From a final compilation of 305 CEMP stars with high- and moderately high-resolution observations in the Galactic halo, they showed that the sample can be broadly divided into CEMP-$s$ (including $\sim$10 CEMP-$r$/$s$ stars) with $A$(C) $> 7.1$, and CEMP-no stars with $A$(C) $\leq 7.1$, enabling classification even with medium-resolution spectra or photometric estimates. The CEMP stars were further subdivided, suggesting multiple progenitor channels. In their Yoon-Beers diagram, stars in the metallicity range $-3.5 \lesssim $ [Fe/H] $ \lesssim -1.5$, with a dispersion of $\sim$0.7\,dex and a peak at $A$(C) $= 7.9$, with weak metallicity dependence, were designated as Group I, comprising mostly CEMP-$s$ stars. The CEMP-no stars were divided into two morphological groups. CEMP-no stars with [Fe/H] $\lesssim -2.5$ were classified into Group II, and exhibit the lowest $A$(C) and a clear correlation with [Fe/H]. Group III stars, the most metal-poor CEMP-no stars ([Fe/H] $< -3.5$), have $A$(C) $\sim 6.8$ and no clear trend with [Fe/H]. 

Incorporating about 50 stars from MW dwarf spheroidal (dSph) and ultra-faint dwarf galaxies (UFDs), \citet{Yoon2019} proposed possible formation pathways: Group I CEMP-$s$ stars likely formed via mass accretion from AGB companions enriched in C and Ba; Group II CEMP-no stars may have originated in dSph clouds cooled by Mg- and Si-rich grains from core-collapse supernovae (CCSNe); and Group III CEMP-no stars possibly formed in UFDs enriched by C from faint SNe and/or spinstars \citep{Chiaki2017}. A subset of the Group I CEMP-no stars may have acquired C via binary mass transfer. However, their low Ba abundances remain unexplained, raising the possibility of a shared origin with Group III stars, an idea that requires further study. On this point, \citet{Rossi2023}, using the semi-analytical data-calibrated model of \citet{Rossi2021}, suggested that Group III CEMP-no stars might also originate from AGB stellar winds without invoking binarity. 

Most recently, \citet[][hereafter L25]{Lee2025} analyzed about 3.2 million medium-resolution ($R \sim 1800$) spectra from the Sloan Digital Sky Survey \citep[SDSS;][]{York2000} and the Large Sky Area Multi-Object Fiber Spectroscopic Telescope \citep[LAMOST;][]{Cui2012}. Based on Gaussian Mixture Modeling (GMM) of the $A$(C) distribution for their SDSS/LAMOST CEMP sample, they suggested a new morphological group of CEMP stars, designated as Group IV. These stars exhibit $A$(C) values comparable to Group I stars but [Fe/H] as low as Group III stars, consisting of a roughly even mix of CEMP-$s$ and CEMP-no stars and exhibiting a binary fraction of $\sim$30–40\%. The authors suggested that these stars might have formed in metal-poor environments with normal or mildly enhanced C, and later acquired additional C via mass transfer from an AGB companion, resulting in their high $A$(C).

The subclasses of CEMP stars can now be examined with much larger samples based on recent large-scale photometric surveys. The multi-band systems (including both narrow/medium-band and broad-band filters) employed in these surveys enable the estimation of stellar metallicities and several elemental abundances, including C and Mg. For instance, \citet{Huang2024} and Huang et al.\ (submitted) re-calibrated the third data release (DR3) of the Javalambre Photometric Local Universe Survey \citep[J-PLUS;][]{Cenarro2019} and DR4 of the Southern Photometric Local Universe Survey \citep[S-PLUS;][]{MendesdeOliveira2019} to improve photometric zero-point accuracy across all filters, and derived stellar parameters such as effective temperature ($T_{\rm eff}$), surface gravity ($\log~g$), [Fe/H], [C/Fe], [Mg/Fe], and [$\alpha$/Fe] for approximately 10.5 million stars. 

Using the combined J-PLUS/S-PLUS dataset of approximately 6.40 million stars, after applying data-quality cuts that improve the precision of the abundance estimates, and other selection criteria, we identify about 104,900 candidate CEMP stars and directly compare it with the spectroscopic study of L25. Through examination of the distribution of $A$(C)$_{\rm c}$ (corrected for evolutionary effects from \citealt{Placco2014c}) for the CEMP candidates as a function of [Fe/H], and applying GMM, we statistically determine multiple CEMP subgroups. Using the threshold $A$(C)$_{\rm c} = 7.15$ to separate CEMP-no and CEMP-$s$ stars, we then explore changes in their relative fractions and cumulative frequencies. We also investigate how the frequencies of these subgroups vary with spatial position in the MW, including the outer halo, inner halo, and thick-disk regions, and compare our results with those based on medium-resolution SDSS spectroscopy from \citet{Lee2017} and \citet{Lee2019}. Leveraging this large photometric sample, the study provides an independent validation of spectroscopic-based results and offers a broader view of the Galactic evolutionary history inferred from the distribution of CEMP stars.

The remainder of this paper is organized as follows. Section~\ref{sec:dataandmethods} describes the J-PLUS and S-PLUS data used in this study. In Section~\ref{sec:results}, we present the distribution of the corrected [C/Fe] abundance ratio, [C/Fe]$_{\rm c}$, and the corrected absolute carbon abundances, $A$(C)$_{\rm c}$, as a function of [Fe/H], for the combined dataset. We also identify the photometrically classified CEMP-$s$ and CEMP-no candidates, classified as Groups I to IV in the Yoon–Beers diagram, along with the variation in the fractions of CEMP-$s$ and CEMP-no stars across different [Fe/H] ranges. In Section~\ref{sec:discussion}, we discuss the spatial distributions of CEMP candidates in the various regions of the Galaxy. We summarize our findings in Section~\ref{sec:summary}.

\section{Data}\label{sec:dataandmethods}

\begin{figure}
    \centering
    \includegraphics[width=\linewidth]{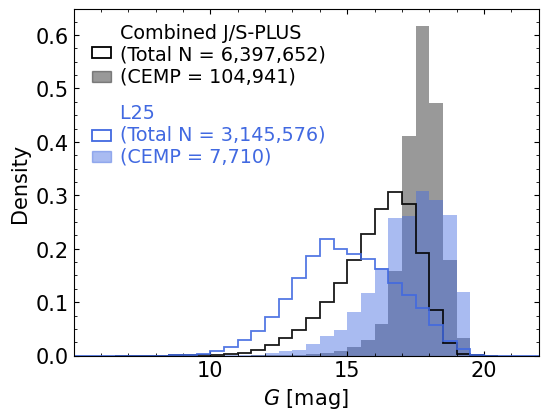}
    \caption{Distribution of $G$ magnitudes for the combined J/S-PLUS sample and the L25 SDSS and LAMOST samples. Unfilled black and blue histograms represent the total J/S-PLUS and L25 samples, respectively, while filled histograms show the distributions of the CEMP stars in the J/S-PLUS and L25 samples. The number of stars in each sample is indicated in the top-left corner.}
    \label{Fig:Gmag}
\end{figure}

J-PLUS is a survey aimed at understanding the star formation history of the local Universe as one of its primary goals. Observing the Northern sky, its multi-filter system enables the estimation of stellar parameters (e.g., $T_\mathrm{eff}$, $\log g$, and [Fe/H]) and chemical-abundance ratios such as [C/Fe] and [Mg/Fe], for MW stars, including the disk and halo populations. Observations are conducted using a 2 deg$^2$ field of view camera, mounted on the 83 cm Javalambre Auxiliary Survey Telescope (T80) at the Observatorio Astrof\'isico de Javalambre \citep[OAJ;][]{Cenarro2014}. The survey utilizes five broad-band filters ($u_{JAVA}, g, r, i, z$) in addition to seven narrow- and medium-band filters ($J0378, J0395, J0410, J0430, J0515, J0660, J0861$), targeting spectral features such as \ion{O}{2}, \ion{Ca}{2}~H\&K, H$\delta$, the CH-$G$ band, \ion{Mg}{1}~b~triplet, H$\alpha$, and the \ion{Ca}{2}~triplet. From the planned footprint of approximately 8,500 deg$^2$, the DR3 covers 3,192 deg$^2$ observed between 2015 and 2022, reaching limiting magnitudes between 20.8 and 21.8 mag \citep{Lopez-Sanjuan2024}. This release includes approximately 47.4 million detected sources.

With the initial internal calibration of J-PLUS, \citet{Lopez-Sanjuan2024} sought to obtain a homogenized photometric solution across the wide-field observations spanning multiple filters by employing the stellar-locus method \citep[SL;][]{Lopez-Sanjuan2019}, with synthetic photometry derived from \Gaia\ DR3 BP/RP (XP) spectra \citep{Carrasco2021, GaiaCollaboration2023}, and Pan-STARRS \citep[PS1;][]{Chambers2016} DR1 photometry. Subsequently, \citet{Xiao2023} performed an additional recalibration using improved \Gaia\ XP spectra and the stellar color-regression method \citep[SCR;][]{Yuan2015}, achieving zero-point precisions on the order of 1–5 mmag. \citet{Huang2024} utilized this re-calibrated data set together with \Gaia\ EDR3 \citep{GaiaCollaboration2021} and high-resolution spectroscopic studies to estimate chemical abundance ratios, including [Fe/H], [C/Fe], [Mg/Fe], and [$\alpha$/Fe], for approximately 5.0  million stars, using a kernel principal component \citep[KPCA;][]{Scholkopf1998} analysis. The precision of these estimates is quantified as 0.10–0.20\,dex for [Fe/H] and [C/Fe] and approximately 0.05\,dex for [Mg/Fe] and [$\alpha$/Fe]. 

\begin{figure*}
    \centering
    \includegraphics[width=\linewidth]{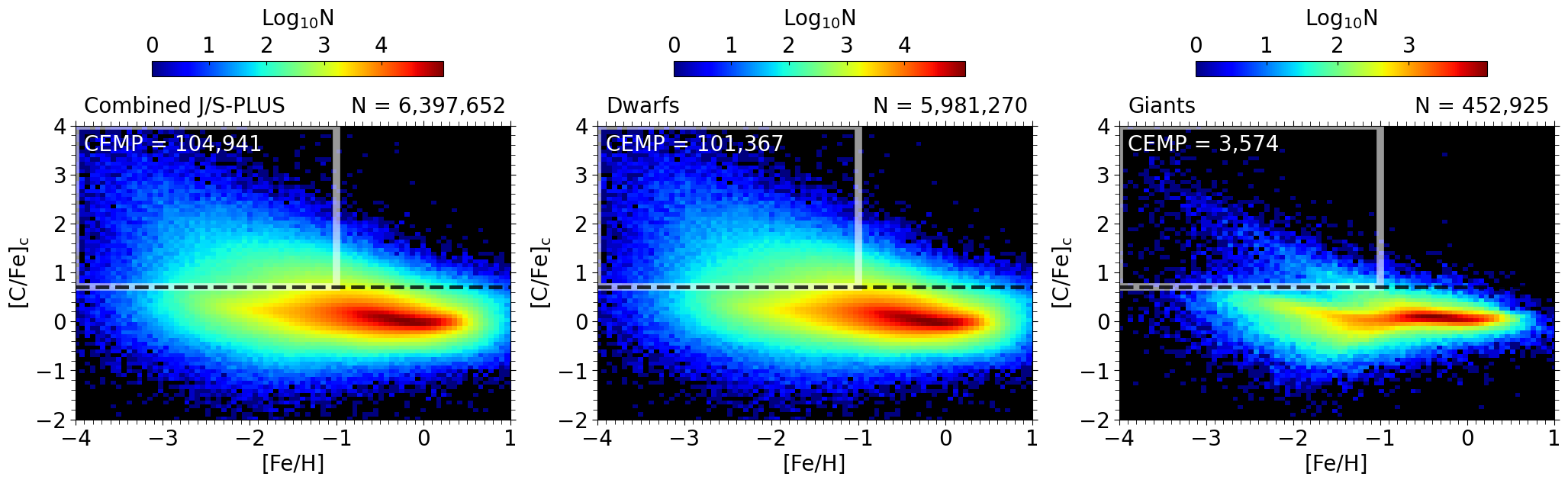}
    \caption{Distribution of [C/Fe]$_{\rm c}$, as a function of [Fe/H], for the combined J/S-PLUS sample. From left to right are all the samples in this study, the dwarf stars, and the giant stars, respectively. The total number of stars is shown in the top-right, and the number density is color-coded on a logarithmic scale. The dashed-black line marks [C/Fe]$_{\rm c}$\ $= +0.7$, and the white box indicates the CEMP candidate region. The number of CEMP candidates is shown in the top-left corner.}
    \label{Fig:fehcfec}
\end{figure*}

S-PLUS is a survey of the Southern sky conducted at the NSF Cerro Tololo Inter-American Observatory using an identical telescope, camera, and filter system as J-PLUS. S-PLUS DR4 \citep{Herpich2024}, covering approximately 3,023 deg$^2$ out of the about 9,300 deg$^2$ target, observed between 2016 and 2021, contains a total of roughly 5.5 million stars. Huang et al.\ (submitted) provide a derivation of stellar-abundance estimates similar to \citet{Huang2024}, based on this data set, for about 5.1 million stars. 

From the J-PLUS and S-PLUS datasets, containing about 10.2 million stars, we selected those with values of [Fe/H]$_\texttt{flg}$ and [C/Fe]$_\texttt{flg}$ greater than 0.85, resulting in approximately 7.5 million stars with reliable abundance estimates. [Fe/H]$_\texttt{flg}$ and [C/Fe]$_\texttt{flg}$ are flags, or data-quality indicators, proposed by Huang et al., which assign a maximum kernel value of 1 when the color indices of the training and target samples used in KPCA are identical. To ensure sample homogeneity, we further excluded stars likely to be members of globular clusters, based on the catalog of \citet{Harris2010}.
In addition, we excluded binaries identified from the color–magnitude diagram by \citet[][submitted]{Huang2024} and further removed stars that may be unresolved binary systems by applying a Gaia renormalized unit weight error cut of RUWE $<$ 1.4. We then imposed a temperature cut of $4,000$ K $\leq T_{\rm eff} \leq 6,700$ K to match the L25 sample and allow a direct comparison with its medium-resolution spectroscopic results. These selections result in a final sample of approximately 6.40 million stars, hereafter referred to as the combined J/S-PLUS sample. 

The Gaia $G$-magnitude distributions of the combined J/S-PLUS and L25 sample are shown in Figure~\ref{Fig:Gmag}. Unfilled black and blue histograms represent the total J/S-PLUS ($\langle G \rangle = 16.0$) and L25 ($\langle G \rangle = 15.0$) samples, respectively, while filled histograms show the distributions of CEMP stars in J/S-PLUS ($\langle G \rangle  = 17.7$) and L25 ($\langle G \rangle = 17.3$). For the total samples, the $G$-magnitude ranges are similar, although the L25 spectroscopic sample includes a larger fraction of bright stars, whereas our photometric sample contains many more faint stars. However, the brightness distributions of the CEMP stars are more comparable in terms of mean and overall shape. 

In this paper, we use [Fe/H] or metallicity to refer to the [Fe/H] estimates derived from the combined sample. The values of [C/Fe] and $A$(C) have been corrected following the evolutionary models of \citet{Placco2014c}, which account for carbon depletion caused by internal mixing during the RGB phase \citep[e.g.,][]{Nguyen2025}, providing [C/Fe]$_{\rm c}$ and $A$(C)$_{\rm c}$ for a more accurate identification and interpretation of CEMP subclasses. In the Appendix, We discuss a comparison of the photometric [Fe/H] and [C/Fe] estimates for CEMP candidates with the spectroscopic measurements of L25.

\begin{figure}
    \centering
    \includegraphics[width=\linewidth]{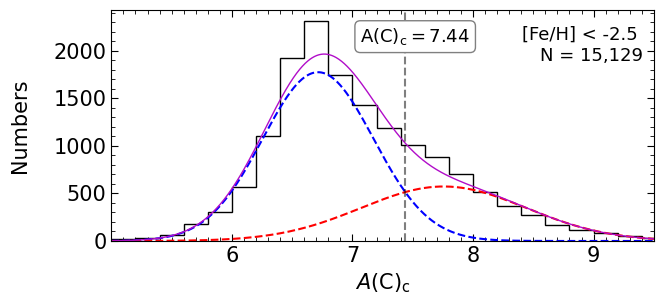}
    \caption{The $A$(C)$_{\rm c}$ distribution of the combined J/S-PLUS CEMP sample with [Fe/H] $< -2.5$, after applying flag cuts of 0.85. The sample size is indicated in the upper-right corner. The dashed-blue and dashed-red curves represent the GMM fits for CEMP-$s$ and CEMP-no candidates, respectively, while the solid-purple curve indicates the sum of the Gaussian fits. The intersection of the two Gaussians is marked by a vertical dashed-black line at $A$(C)$_{\rm c} = 7.44$.} 
    \label{Fig:gmm_feh25}
\end{figure}

\begin{figure*}
    \centering
    \includegraphics[width=\linewidth]{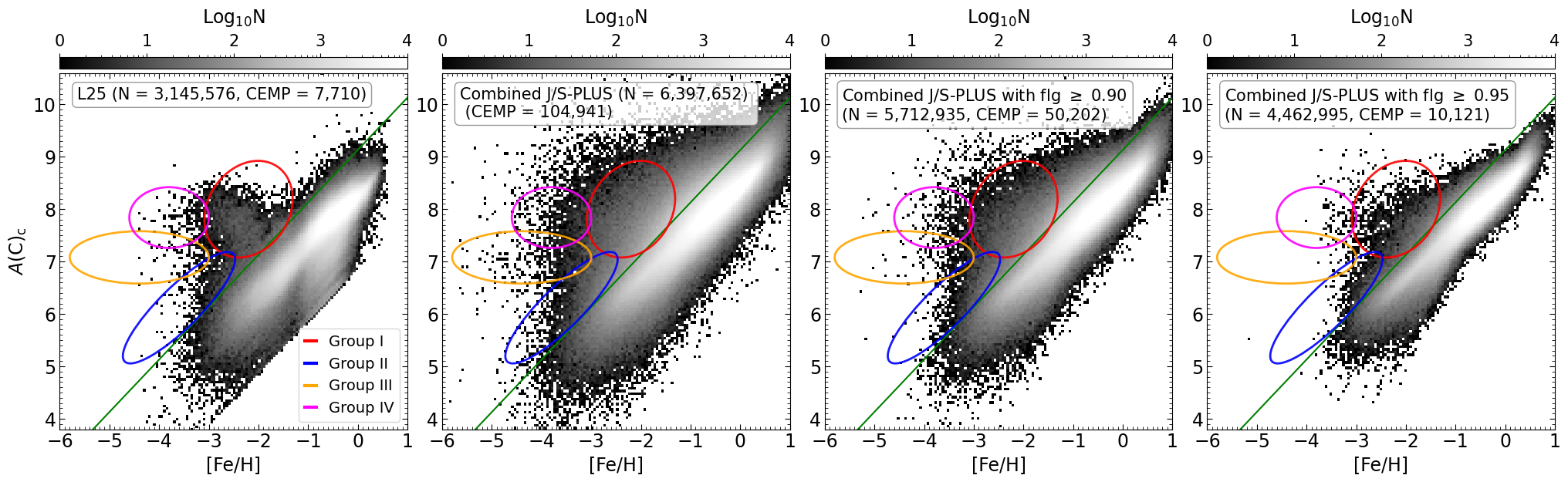}
    \caption{Distribution of $A$(C)$_{\rm c}$, as a function of [Fe/H], for the combined J/S-PLUS sample. The solid-green line represents [C/Fe]$_{\rm c}= +0.7$. The red, blue, orange, and magenta ellipses represent subgroups of CEMP stars, corresponding to Groups I, II, III, and IV, respectively. The left-most panel shows the SDSS and LAMOST DR6 sample from L25, and the second through fourth panels show the J/S-PLUS sample with data-quality flag cuts of 0.85, 0.90, and 0.95, respectively. The total number of stars in each sample is indicated at the top of each panel.}
    \label{Fig:yoonbeersdiagram}
\end{figure*}

\begin{figure}
    \centering
    \includegraphics[width=\linewidth]{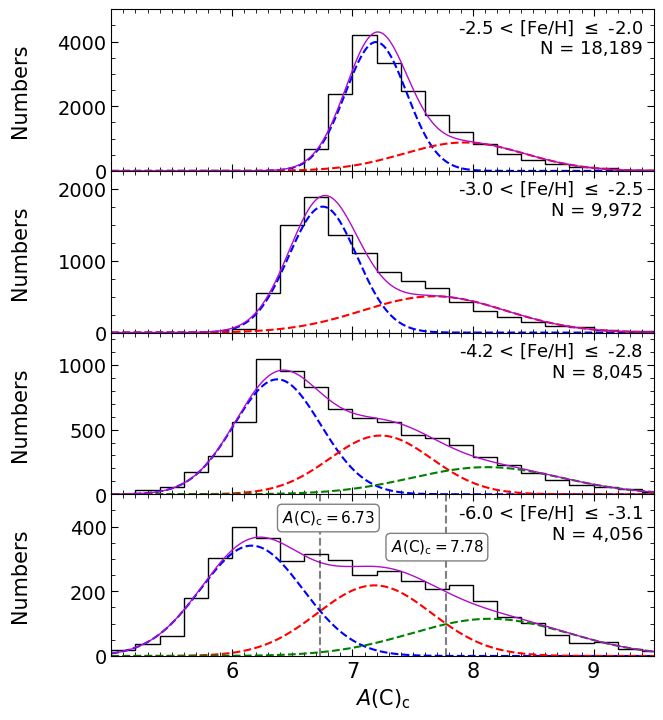}
    \caption{The $A$(C)$_{\rm c}$ distribution of the combined J/S-PLUS CEMP sample with [Fe/H] $\leq -2.0$, after applying flag cuts of 0.85. The [Fe/H] ranges for each panel are displayed in the top-right corner along with the sample size. The dashed-blue, dashed-red, and dashed-green curves represent the GMM fits for each component, while the solid-purple curve denotes the fit for the sum of all components. The intersections of the Gaussians, representing the separation between Groups II, III, and IV, are marked by vertical dashed-black lines at $A$(C)$_{\rm c} = $ 6.73 and 7.78.}
    \label{Fig:gmm}
\end{figure}

\begin{figure*}
    \centering
    \includegraphics[width=\linewidth]{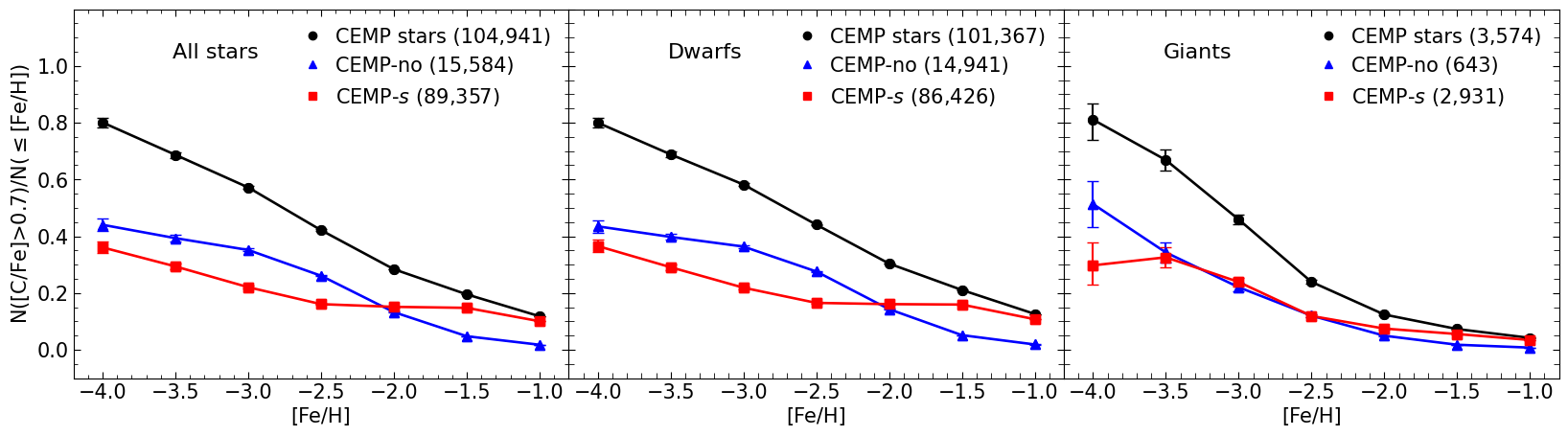}
    \caption{Cumulative frequencies of CEMP stars for each [Fe/H] bin. Solid-black lines with circular markers show the fraction of CEMP candidates among the combined J/S-PLUS sample. Each marker represents the cumulative fraction of CEMP stars among stars with [Fe/H] values less than or equal to the corresponding bin. When dividing the CEMP candidates into CEMP-no and CEMP-$s$ types based on $A$(C)$_{\rm c} = +7.15$, their respective fractions are shown with triangle markers and solid-blue lines (CEMP-no), and square markers with solid-red lines (CEMP-$s$). The total number of stars in each sample is indicated in the top-right corner. Errors are calculated using 68\% Wilson score intervals for bins with fewer than 40 stars, and the normal approximation for bins with 40 or more.  Note that the Group IV stars are included in this sample.}
    \label{Fig:cempfrac}
\end{figure*}

\begin{figure*}
    \centering
    \includegraphics[width=\linewidth]{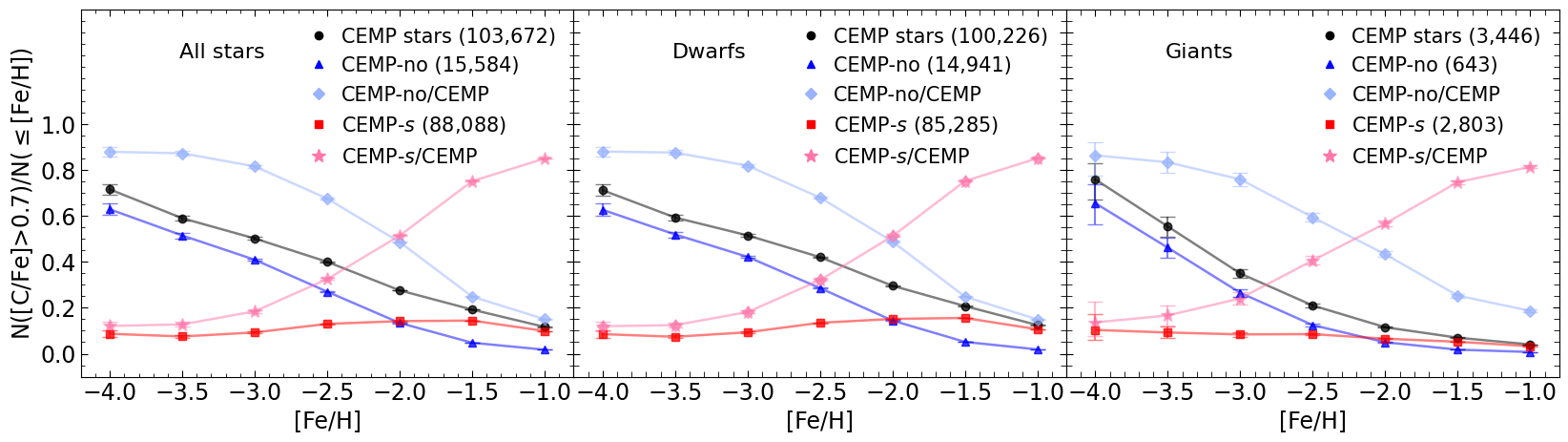}
    \caption{Similar to Figure~\ref{Fig:cempfrac}, but excluding Group IV stars. The sky-blue diamonds and solid line indicate the fraction of CEMP-no stars among CEMP stars, while the pink stars and solid line indicate the fraction of CEMP-$s$ stars among CEMP stars.}
    \label{Fig:cempfrac_wo_groupiv}
\end{figure*}

\section{Results}\label{sec:results}

Here we examine the variation of [C/Fe]$_{\rm c}$ and $A$(C)$_{\rm c}$, as a function of [Fe/H], for the identified candidate CEMP stars. 

\subsection{Metallicity vs. Carbonicity Distribution}\label{sec:fehcfeacc}

Figure~\ref{Fig:fehcfec} shows the distribution of [C/Fe]$_{\rm c}$ (carbonicity), as a function of [Fe/H], for the combined J/S-PLUS sample. From left to right, the panels show the distributions for the entire sample, dwarf stars, and giant stars, respectively, where dwarfs and giants are classified following L25, using $\log g > 3.5$ for dwarfs and $\log g \leq 3.5$ for giants. The number density is color-coded on a logarithmic scale. The dashed-black line shows [C/Fe]$_{\rm c} = +0.7$, and the white box in the top-left corner marks the CEMP candidate stars. All three panels of this figure clearly show an increase in [C/Fe]$_{\rm c}$ with decreasing [Fe/H], as reported in previous studies.  We identified 104,941 CEMP candidates; 96.6\% of this sample are dwarfs and 3.4\% are giants.

According to \citet{Yoon2016}, CEMP stars with low values of $A$(C)$_{\rm c}$ are generally associated with the CEMP-no class, while those with high values of $A$(C)$_{\rm c}$ correspond to the CEMP-$s$ class. Based on this, we investigated whether CEMP stars can be separated into subgroups using photometric estimates of $A$(C)$_{\rm c}$. Figure~\ref{Fig:gmm_feh25} shows the results of applying a two-component GMM in the $A$(C)$_{\rm c}$ space to CEMP candidates in the combined J/S-PLUS sample with [Fe/H] $< -2.5$, using a flag cut of 0.85. In the figure, the dashed-blue and dashed-red curves represent the low and high $A$(C)$_{\rm c}$ groups, which can be considered to correspond to the CEMP-no and CEMP-$s$ classes \citep{Yoon2016}. Note that some stars near the boundary may overlap between the two populations.  The solid-purple curve shows the sum of the two Gaussian components.

The vertical dashed-black line marks the division at $A$(C)$_{\rm c} = 7.44$, approximately 0.3\,dex higher than the values of 7.1 and 7.15 reported by \citet{Yoon2016} and L25, respectively. To identify the origin of this difference, similar analyses were performed for dwarfs and giants, resulting in division values of 7.45 and 7.01, respectively. Furthermore, fitting the GMM to dwarf subsamples limited to stars with $g \leq$ 18, 17, and 16 yields division values of 7.38, 7.29, and 7.18, respectively. This indicates that the higher division relative to the spectroscopic value is primarily driven by the fainter dwarf stars in the photometric sample.  However, since the goal of this study is to independently assess the L25 results using photometric data, we chose to apply the same criterion as in the previous study. Consequently, we adopt the proposed L25 value of 7.15 to separate CEMP-no and CEMP-$s$ stars.

We now examine the Yoon–Beers diagram, which shows the distribution of $A$(C)$_{\rm c}$ as a function of [Fe/H]. Figure~\ref{Fig:yoonbeersdiagram} presents the combined J/S-PLUS sample, where the solid-green line denotes [C/Fe]$_{\rm c} = +0.7$. The red, blue, and orange ellipses represent CEMP Groups I, II, and III as defined by \citet{Yoon2016}, while the magenta ellipse corresponds to Group IV, first identified by L25. The left panel shows how L25 divided CEMP stars into groups using their medium-resolution spectroscopic sample from SDSS \citep{Yanny2009, Rockosi2022, Dawson2013, Blanton2017} and LAMOST DR6, while the three panels on the right illustrate how the much larger photometric sample maps onto the spectroscopic-based grouping. The second, third, and fourth panels correspond to photometric [Fe/H] and [C/Fe]$_{\rm c}$ flag values $\geq$ 0.85, 0.90, and 0.95, respectively. The photometric results exhibit features that are distinct from the L25 results shown in the left-most panel, with the highest-quality sample (flag $\geq 0.95$) in the right-most panel exhibiting pronounced groupings at [C/Fe]$_{\rm c} \ge +0.7$ due to the reduced number of Group I stars, producing a pattern closer to that of the left-most panel. Additionally, the lack of [C/Fe]$_{\rm c}$ values below $–1.6$ in the spectroscopic sample is less pronounced in the photometric sample, where values extend only down to about $–1.2$, likely due to the lower sensitivity of the photometric estimates, which can result in slightly higher $A$(C)$_{\rm c}$ for the CEMP subgroups identified photometrically.

To examine whether the new Group IV stars with very high $A$(C)$_{\rm c}$ values identified by L25 at the lowest metallicities (as well as the other subgroups) can be identified using photometric data, we performed GMM fitting to the $A$(C)$_{\rm c}$ distributions of our sample stars with quality flags $\geq$ 0.85 over selected ranges of [Fe/H]. Following Figure 5 of L25, we first consider the upper two panels of Figure~\ref{Fig:gmm}, where two Gaussian components are fitted over $-3.0 <$ [Fe/H] $\leq -2.0$. The Gaussian associated with high values of $A$(C)$_{\rm c}$ (dashed-red curve) exhibits a significantly lower peak than that for low values of $A$(C)$_{\rm c}$ (dashed-blue curve).  This pattern remains when subjected to stricter flag cuts. There is a contrast in this pattern with the results of L25, where the two components have comparable peak heights. This likely occurs because the photometrically defined Groups II–IV contain substantially more stars than Group I stars.

For the lower two panels, restricted to stars with $-6.0 <$ [Fe/H] $\leq -2.8$, we compared models with two and three Gaussian components using the Akaike Information Criterion (AIC) and Bayesian Information Criterion (BIC). L25 reported that, for the [Fe/H] ranges corresponding to the upper two panels, the two-component model is preferred, while for the lower two panels, the three-component model is preferred. In contrast, for the combined J/S-PLUS sample, our internal checks show that across the [Fe/H] ranges of all four panels, the three-component GMM consistently yields lower AIC and BIC values and is thus preferred; this result remains unchanged when the quality-flag cuts are increased. However, in the range $-3.0 <$ [Fe/H] $\leq -2.0$, our primary goal is to separate the stars of Group I and Group II, while for $-6.0 <$ [Fe/H] $\leq -2.8$, it is to compare directly with L25 to see whether Groups II–IV exhibit three components. For this reason, only two GMM components are shown in the upper two panels, whereas the lower two panels present three components, including a new Gaussian at the highest $A$(C)$_{\rm c}$ region.

For [Fe/H] $\leq -3.1$, the intersections of the three Gaussian components in the bottom panel of Figure~\ref{Fig:gmm} occur at $A$(C)$_{\rm c} = 6.73$ and 7.78.  Compared to the L25 values of 6.74 and 7.39, the first intersection value in the photometric data is similar, but the second value is approximately 0.39\,dex higher. This is because the mean values of the $A$(C)$_{\rm c}$ of the second and third Gaussian curves in the photometric sample are slightly higher; the means and standard deviations of the three curves (from left to right) are 6.16 $\pm$ 0.18, 7.18 $\pm$ 0.22, and 8.14 $\pm$ 0.43, respectively. 

Considering that the lowest extreme of the photometric carbon abundances in Figure~\ref{Fig:yoonbeersdiagram} is about 0.4\,dex higher than in the spectroscopic sample, adjusting the photometric $A$(C)$_{\rm c}$ values and performing the GMM fitting shifts the curves to 6.20 $\pm$ 0.20, 7.14 $\pm$ 0.21, and 7.98 $\pm$ 0.41, while the intersections remain at 6.73 and 7.79, with the second intersection still higher than in L25.\footnote{Following this adjustment, Figure~\ref{Fig:gmm_feh25} gives a value of 7.44.} In other words, even after correcting for the difference in carbon abundances between spectroscopic and photometric samples to obtain three curves with similar means, the second intersection -- relevant for defining Groups III and IV -- remains elevated in the photometric data, likely due to the photometric Group II sample having more stars than the Group IV sample.

\subsection{Frequencies of CEMP Stars}\label{sec:freqcemp}

Following L25, we divide stars according to their $A$(C)$_{\rm c}$ values, classifying those with $A$(C)$_{\rm c} \leq 7.15$ as CEMP-no and those with $A$(C)$_{\rm c} > 7.15$ as CEMP-$s$. We then examine the cumulative frequencies of total CEMP, CEMP-no, and CEMP-$s$ stars among all stars with [Fe/H] below each cutoff value, as shown in Figure~\ref{Fig:cempfrac}. Solid-black lines with filled circles represent the cumulative fraction of CEMP candidates in each [Fe/H] bin, while the blue triangles and red squares indicate the contributions from CEMP-no and CEMP-$s$ stars, respectively. For bins containing fewer than 40 stars, we estimated binomial confidence intervals using the 68\% (1$\sigma$) Wilson score interval \citep{Wilson1927}, following \citet{Brown2001}, \citet{Yoon2018}, and L25; for bins with 40 or more stars, we adopted the normal approximation.

In the left panel of Figure~\ref{Fig:cempfrac}, the CEMP frequency increases from 12\% to 80\% over the range $-4.0 \leq$ [Fe/H] $\leq -1.0$. The CEMP-no fraction increases from about 2\% to 44\%, while the CEMP-$s$ fraction increases more gradually, from 10\% to 36\%. For [Fe/H] $\leq -3.5$, both fractions are shown, with respective uncertainties of roughly 1--2\%. Compared to the L25 results, the overall trend is similar. However, in our results, the fraction of CEMP-no stars becomes higher starting from [Fe/H] $\leq -2.0$, while most CEMP stars are dominated by CEMP-$s$ at higher metallicities. Furthermore, taking into account the uncertainties in the analysis of L25 at [Fe/H] $\leq -4.0$, our CEMP-no results are higher than their low-resolution spectroscopic results, but lower than their high-resolution results, while our CEMP-$s$ results are lower than the high-resolution results and consistent with the uncertainties of the low-resolution results. 

We examined the CEMP frequencies in more detail by separating stars into dwarfs and giants. In the middle panel, the dwarfs show fraction distributions and uncertainties similar to those of the full sample, as expected, which is likely due to the photometric sample including many fainter dwarfs across the entire sky. In contrast, the pattern for giants in the right panel is somewhat different: the fraction of CEMP-$s$ stars decreases slightly (from about 33\% at [Fe/H] $\leq -3.5$ to 30\% at [Fe/H] $\leq -4.0$), while the fraction of CEMP-no stars rises from roughly 34\% to 51\% at the lowest metallicities, with uncertainties of approximately 4\% and 8\% for both fractions in the two lowest-metallicity bins. These trends indicate that CEMP-no stars dominate among the most metal-poor CEMP giants.

\section{Discussion}\label{sec:discussion}

The cumulative frequencies shown in Figure~\ref{Fig:cempfrac} are the fractions of CEMP, CEMP-no, and CEMP-$s$ stars relative to \textit{all} stars. However, the frequencies presented above (and also in L25) stand in contrast to previous work (e.g., \citealt{Yoon2018}), in particular, regarding the behavior of the CEMP-no and CEMP-$s$ stars. Note that previous samples of CEMP stars did not identify numerous stars that we now recognize belong to Group IV, those with very low [Fe/H] but high $A$(C)$_{\rm c}$.  It is thus important to consider their impact on the frequency distributions of stars classified as CEMP-no and CEMP-$s$. We also consider a related question, first raised by L25 for their spectroscopic sample:  Among stars classified as CEMP, how do the fractions of CEMP-no and CEMP-$s$ stars compare with each other as a function of metallicity?

Figure~\ref{Fig:cempfrac_wo_groupiv} shows the cumulative CEMP frequency after excluding the 1,269 Group IV stars with [Fe/H] $\leq -3.1$ and $A$(C)$_{\rm c} > +7.39$, the criteria used by L25, allowing for a direct comparison with Figure 7 of L25. The overall fraction for all CEMP stars is very similar to the frequencies obtained before removing Group IV, with only slightly lower fractions appearing at [Fe/H] $\leq -2.5$. However, compared to Figure~\ref{Fig:cempfrac}, the most noticeable difference is the reduced fraction of CEMP-$s$ stars in all [Fe/H] ranges, with all values falling below 15\%. This trend is also seen in the CEMP frequency patterns of both dwarfs and giants in the middle and right panels. It is clear that the presence of the Group IV stars has a significant influence on the cumulative CEMP frequency diagram.  When Group IV stars are removed, the trends are now consistent with those reported by L25 from both their low- and high-resolution spectroscopic samples, as well as with the trends reported by \citet{Yoon2018} based on the AAOmega Evolution of Galactic Structure (AEGIS) spectroscopic survey.

Figure~\ref{Fig:cempfrac_wo_groupiv} also shows the fractions of CEMP-no and CEMP-$s$ stars among all stars classified as CEMP. From inspection, the fraction of CEMP-no stars among CEMP stars, indicated by the sky-blue diamonds, is quite low ($\sim$ 15\%) at high metallicity, but rises sharply toward lower metallicities, reaching $\sim$ 88\%. In contrast, the fraction of CEMP-$s$ stars, indicated by the pink stars, is high ($\sim$ 85\%) at high metallicity, but declines rapidly with decreasing metallicity, to $\sim$ 12\%. The cross-over between the relative dominance of the two subclasses of CEMP stars occurs at [Fe/H] = $-2.0$.  

\begin{figure*}
    \centering
    \includegraphics[width=\linewidth]{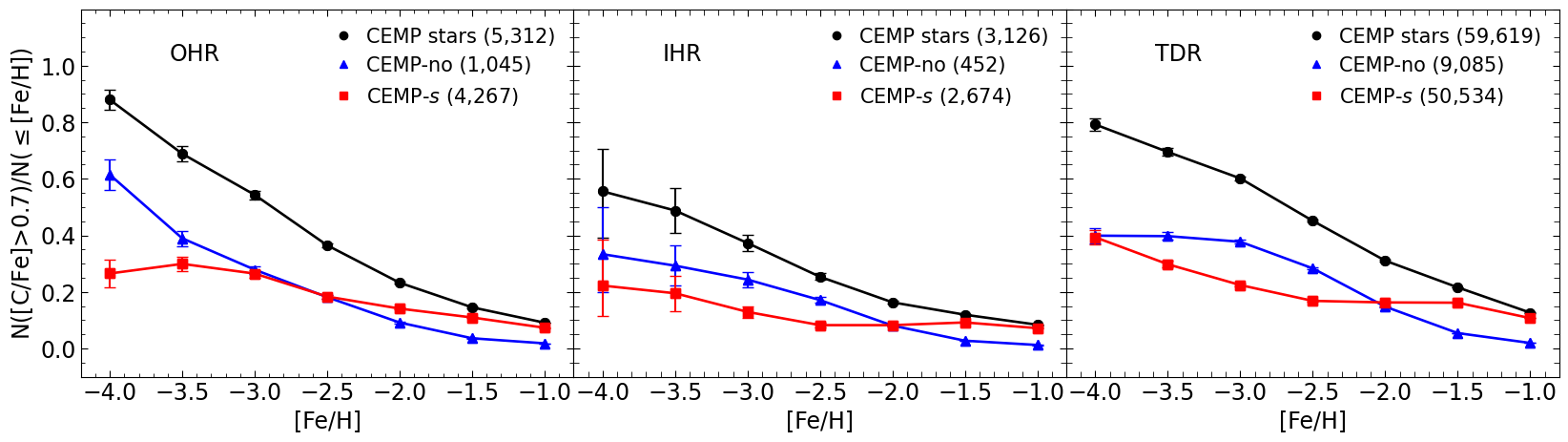}
    \caption{Similar to Figure~\ref{Fig:cempfrac}, but showing the cumulative frequency of CEMP stars in each [Fe/H] bin across different regions of the MW. From left to right, the panels represent the fractions in the outer-halo region (OHR), inner-halo region (IHR), and thick-disk region (TDR), respectively.  Note that these samples include the Group-IV stars.}
    \label{Fig:cempfrac_byregions}
\end{figure*}

\begin{figure*}
    \centering
    \includegraphics[width=\linewidth]{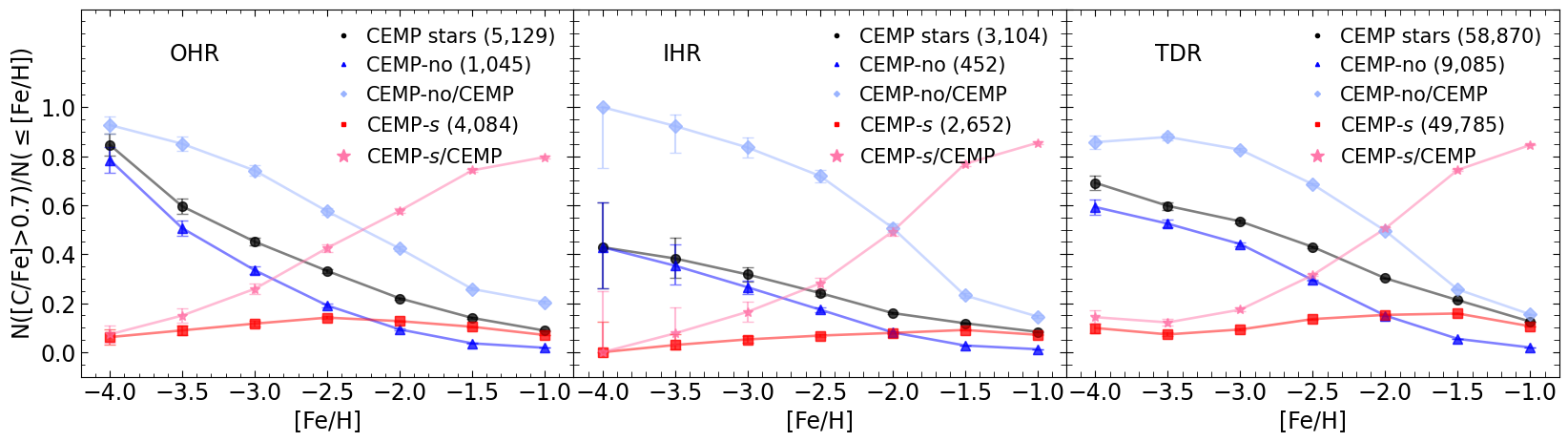}
    \caption{Similar to Figure~\ref{Fig:cempfrac_byregions}, but excluding Group IV stars.}
    \label{Fig:cempfrac_byregions_wogroupiv}
\end{figure*}

L25 suggested that the strong influence of the presence of Group IV stars in the cumulative frequency diagrams could arise because some Group IV stars, which might actually be CEMP-no, were classified as CEMP-$s$ according to the $A$(C)$_{\rm c} > +7.15$ criterion. They also proposed that these metal-poor, carbon-rich Group IV CEMP stars may have a complex nucleosynthesis history -- forming as second-generation stars in natal clouds with low metallicity, while simultaneously being enriched in carbon through mass transfer from AGB stars (either in binary or non-binary configurations). The overall agreement between L25's low-resolution spectroscopic sample of roughly 3.15 million stars, including about 7,700 CEMP stars, and our photometric sample of about 6.40 million stars, including around 104,900 CEMP candidates, supports the interpretation proposed by L25.

\subsection{CEMP Frequencies in Different Regions of the MW}

Figure~\ref{Fig:cempfrac_byregions} presents the cumulative CEMP frequencies of the combined J/S-PLUS sample in different spatial regions of the MW, including the outer-halo region (OHR), inner-halo region (IHR), and thick-disk region (TDR), following the definitions of \citet{Lee2017}. From left to right, the panels show the OHR ($|Z| > 3$ kpc and $R > 11.5$ kpc), the IHR ($|Z| > 3$ kpc and $R \leq 8.5$ kpc), and the TDR ($|Z| \leq 3$ kpc and $R \leq 8.5$ kpc), where $Z$ is the height above the Galactic plane and $R$ is the distance from the Galactic center projected onto the plane.

Within the TDR, the fraction of CEMP stars ranges from approximately 13\% to 79\%, and the frequencies of both CEMP-no and CEMP-$s$ stars are generally similar to those observed in dwarf stars over the metallicity range [Fe/H] between −3.0 and −1.0, as shown in Figure~\ref{Fig:cempfrac}; indeed, about 99\% of these stars have been confirmed to be dwarf stars. However, the fraction of CEMP-no stars remains constant toward lower metallicities below −3, resulting in nearly equal fractions of the two subgroups at [Fe/H] $\leq -4.0$, each at 40\% with an uncertainty of about 3\%. In the IHR, about 90\% of the stars are dwarf stars, and this region also exhibits a pattern similar to that observed in dwarf stars when regional distinctions are not made, with CEMP-no at 33\% ($^{+17}_{-13}$) and CEMP-$s$ at 22\% ($^{+16}_{-11}$) at the lowest metallicities. However, the overall fraction of CEMP stars is roughly 20\% to 30\% lower than in the TDR and OHR. In contrast, in the OHR, the frequencies resemble those of giant stars in that figure, as giants account for up to about 32\%, with the fraction of CEMP stars rising to approximately 88\%. In addition, the CEMP-$s$ frequency overtakes the CEMP-no frequency starting at [Fe/H] $> -3.0$, unlike in the IHR and TDR. Moreover, at [Fe/H] $\leq -4.0$, the frequency of CEMP-no stars reaches about 61\%, whereas the frequency of CEMP-$s$ stars remains similar to that in the other regions, at 27\% in the OHR, with an uncertainty of 5\% for both.

Figure~\ref{Fig:cempfrac_byregions_wogroupiv} is similar to Figure~\ref{Fig:cempfrac_byregions}, but shows the spatial CEMP cumulative frequencies after excluding Group IV stars. In this figure, the distribution of CEMP-no and CEMP-$s$ stars across regions is more clearly visible. For example, in the TDR, the CEMP-no fraction continues to increase steadily, as in the results without regional distinctions (Figure~\ref{Fig:cempfrac_wo_groupiv}). In the IHR, although the overall fraction of CEMP stars is still lower than in the other regions, CEMP-no fractions among CEMP stars (sky-blue) start to dominate among CEMP stars at [Fe/H] $\leq -2.0$, reaching nearly all CEMP stars at [Fe/H] $\leq −4$. In the OHR, the frequency of CEMP stars, mostly CEMP-no, increases sharply toward the lowest metal-poor regime. In addition, the fractions of CEMP-no and CEMP-$s$ stars among CEMP stars cross at approximately [Fe/H] $= −2.2$ as shown in Figure~\ref{Fig:cempfrac_wo_groupiv} for giant stars.

From the results presented in the two figures, the finding that about 90\% of all stars in the OHR are CEMP stars, and that approximately 70--93\% of these are likely second-generation CEMP-no candidates, indicates that they can serve as prime targets for future high-resolution spectroscopic follow-up observations to determine the abundances of heavy elements and their kinematics, providing crucial insights into the early Galaxy. The predominance of CEMP-no stars in the OHR is also consistent with the results of \citet{Carollo2014}, based on dynamical properties from high-resolution spectroscopy, and \citet{Lee2017}, based on chemical properties from medium-resolution spectroscopy.

In the IHR, at [Fe/H] $\leq -3.5$ and $−4.0$, the numbers of CEMP-no and CEMP-$s$ stars are small (fewer than 40), leading to large uncertainties, and may even reverse the observed CEMP-no to CEMP-$s$ fractions. Nevertheless, the overall frequency of CEMP stars remains lower than in the OHR, and the pattern observed in the OHR -- from a sharp increase in the CEMP-no fraction and a decline in the CEMP-$s$ fraction between [Fe/H] $\leq -3.5$ and $−4.0$ -- is not seen in the IHR (Figure~\ref{Fig:cempfrac_byregions}). This suggests that the two regions have intrinsically different properties, supporting the duality of the Galactic halo reported in previous studies \citep[e.g.,][]{Carollo2007,Carollo2010,Carollo2012,Carollo2014,Lee2017,Lee2019,Yoon2018,CabreraGarcia2024} and implying that the evolutionary history of this substructure differs from that of the OHR.

In the TDR, the relative CEMP-no fraction does not increase toward the lowest metallicities, unlike in the other two regions, suggesting that a substantial portion of the most metal-poor stars are CEMP-$s$ stars in binary systems, in contrast to the higher likelihood of finding CEMP-no stars at the same metallicities in the halo regions. Since the formation of CEMP-$s$ stars, involving mass transfer from their AGB companions, likely required a somewhat longer timescale than that of CEMP-no stars, this provides insight into the relative formation epochs of this region and halo. Conversely, the higher overall fraction of CEMP stars, with CEMP-no stars still accounting for about 40\% of all stars at [Fe/H] $\leq -3.0$, suggests that some stars in this region could have been present since the early stages of the Galaxy. Finally, note that there is likely some contamination near the boundaries of each region.

\section{Summary}\label{sec:summary}

We analyze CEMP candidate stars using photometric estimates of [Fe/H] and [C/Fe] from the J-PLUS and S-PLUS samples provided by \citet{Huang2024} and Huang et al.\ (submitted). From a combined sample of approximately 6.40 million stars selected with data-quality flags $\geq 0.85$ and effective temperatures between 4,000 and 6,700 K, we identified 104,941 CEMP stars (101,367 dwarfs and 3,574 giants). This represents an increase of roughly 70 times over the $\sim$1,540 CEMP stars in \citet{Yoon2018} from the AEGIS survey and about 14 times the 7,710 low-resolution CEMP stars in the most recent spectroscopic analysis of SDSS/LAMOST stars by L25.  In addition, our sample of CEMP candidates is approximately twice as large as the 58,872 CEMP candidates identified from Gaia XP spectra by \citet{Lucey2023}. The Gaia $G$-magnitudes of their sample span from 14 to 19 mag (see their Figure 8).  Although this magnitude range is similar to that of our sample (Figure~\ref{Fig:Gmag}), our study includes a larger number of candidates, likely due to the different criteria used to evaluate the quality of candidates in the two samples.
 
Applying a GMM analysis to the CEMP candidates with [Fe/H] $\leq -2.5$, we identified two components with a divided at $A$(C)$_{\rm c} = 7.44$, slightly higher than the 7.15 division reported by L25, likely due to our inclusion of fainter dwarfs with higher $A$(C)$_{\rm c}$.  We examine and compare the CEMP morphological groups (I–IV) proposed by \citet{Yoon2016} and L25, originally defined from spectroscopic samples, and assess their distinguishability using photometric data. Based on the lower AIC and BIC values obtained for a three-component GMM, we photometrically confirm that the sample consists of three groups in the [Fe/H] $\leq -2.8$ regime, including Group IV stars with extremely low metallicity and high $A$(C)$_{\rm c}$ identified by L25 as well as Groups II and III from \citet{Yoon2016}.

To independently test the results of L25 photometrically, and in order to directly compare with their classification, we adopt the $A$(C)$_{\rm c} = 7.15$ threshold of L25 to divide CEMP-no and CEMP-$s$ stars. We then examine their cumulative frequencies relative to all stars in our sample. We find that the fraction of CEMP stars increases toward lower metallicities, and that in the [Fe/H] $\leq -2.0$ regime, CEMP-no stars outnumber CEMP-$s$ stars, particularly among dwarfs. Among giants, the CEMP-no and CEMP-$s$ fractions remain comparable down to [Fe/H] $\leq -3.5$, but CEMP-no stars dominate at [Fe/H] $\leq -4.0$, consistent with the low-resolution frequency results of L25. When Group IV stars are excluded, the CEMP-no fraction increases to $\sim$88\% at [Fe/H] $\leq -4.0$, a behavior consistent with previous work and similar to that of L25. We also consider the relative fractions of CEMP-no and CEMP-$s$ stars among all CEMP stars, and find that CEMP-no stars are the dominant class at [Fe/H] $< -2.0$, while CEMP-$s$ stars dominate above this metallicity.

Finally, based on our very large photometric sample of CEMP candidates, we examined the frequency of CEMP stars in different regions of the MW, including the OHR, IHR, and TDR, as defined by \citet{Lee2017}. In the TDR and IHR, CEMP-no stars become the dominant subgroup at [Fe/H] $\leq -2.0$. However, in the TDR, the relative fraction of CEMP-no stars shows little further increase below [Fe/H] $\leq -3.0$, suggesting that a significant fraction of the most metal-poor stars in this region are CEMP-$s$ stars. Given that CEMP-$s$ stars are thought to form through mass transfer from the AGB companions and therefore trace longer evolutionary timescales, this may indicate that the TDR formed later than the other two halo regions. Nevertheless, because CEMP-no stars still constitute a non-negligible fraction, some TDR members may also date back to the early stages of the Galaxy.

In contrast, the IHR exhibits an overall lower frequency of CEMP stars than the other regions. Moreover, the small number of stars at [Fe/H] $\leq -3.5$ results in large uncertainties in the fractions of both CEMP-no and CEMP-$s$ stars. However, this pattern, which differs markedly from that of the OHR, supports the duality of the Galactic halo as reported in previous studies, and suggests that the IHR may have experienced a distinct formation history compared to the OHR. On the other hand, the OHR exhibits an increasing contribution from CEMP-no stars below [Fe/H] $\leq -3.0$, with a particularly sharp rise at [Fe/H] $\leq -4.0$. The finding that about 90\% of all stars in this region are CEMP, and that approximately 70--93\% of the CEMP stars are CEMP-no stars in the lowest metallicity regime, suggests that these candidate stars are prime targets for future high-resolution spectroscopic follow-up observations, which will be crucial for improving our understanding of the early evolutionary history of the Milky Way.

Adding to the 87,774 Group I, 15,070 Group II, 828 Group III, and 1,269 Group IV samples identified in this study, the J-PLUS and S-PLUS surveys (both nearing completion and likely to be finished within the next two years) are expected to increase the final number of CEMP candidate stars to on the order of half a million. Analyzing the spectra of photometric CEMP candidates in each subclass or subgroup from upcoming large-scale medium- to high-resolution spectroscopic surveys will enable more precise estimates of iron and carbon abundances, as well as measurements of neutron-capture elements. Such analyses will provide insights into the properties of early-generation stars that contributed to the presently observed diversity of CEMP morphological groups, and will help further constrain the evolutionary history of the Galaxy.

\vspace{2.0cm}
\begin{acknowledgements}

T.C.B. acknowledges partial support from grants PHY 14-30152; Physics Frontier Center/JINA Center for the Evolution of the Elements (JINA-CEE), and OISE-1927130; The International Research Network for Nuclear Astrophysics (IReNA), awarded by the US National Science Foundation, and DE-SC0023128; the Center for Nuclear Astrophysics Across Messengers (CeNAM), awarded by the U.S. Department of Energy, Office of Science, Office of Nuclear Physics. Y.S.L. acknowledges support from the National Research Foundation (NRF) of Korea grant funded by the Ministry of Science and ICT (RS-2024-00333766). The work of V.M.P. is supported by NOIRLab, which is managed by the Association of Universities for Research in Astronomy (AURA) under a cooperative agreement with the U.S. National Science Foundation. Y.H. acknowledges the support from the National Science Foundation of China (NSFC grant No. 12422303). J.C.G acknowledges support from the GEM Fellowship supported in part by the National GEM Consortium and partial support by the U.S. DOE through contract numbers DE-FG02-95-ER40934 and LA22-ML-DE-FOA-2440. 
Portions of this text were written with the assistance of ChatGPT \citep*{openai}, especially in regard to English usage and grammar.

\end{acknowledgements}

\begin{appendix}

\setcounter{figure}{0}
\setcounter{table}{0}
\renewcommand{\thetable}{A\arabic{table}}
\renewcommand{\thefigure}{A\arabic{figure}}

Table~\ref{tab:colsdescription} provides descriptions of the parameters for the 104,941 candidate stars from the J/S-PLUS combined sample, including Gaia IDs, positions, magnitudes, colors, distance estimates, extinction values, effective temperatures, surface gravities, stellar types (with data quality labels; see below), abundances, and assigned morphological groups. The CEMP morphological group assignment is based on the criteria of \citet{Zepeda2023}, with slight modifications to ensure consistency with L25: Group I comprises 87,774 stars with [Fe/H] $> -3.1$ and $A$(C)$_{\rm c} > 7.15$; Group II comprises 15,070 stars with $A$(C)$_{\rm c} < 7.15$ and [C/Fe]$_{\rm c} < 1.77$; Group IV comprises 1,269 stars that satisfy $A$(C)$_{\rm c} > 7.39$ and [Fe/H] $\leq -3.1$; and Group III comprises the remaining 828 stars. 

Table~\ref{tab:colsdescription} also lists stars in common with previous spectroscopic studies, such as L25, JINAbase \citep{Abohalima2018}\footnote{https://jinabase.pythonanywhere.com/}, the Stellar Abundances for Galactic Archaeology \citep[SAGA,][]{Suda2008} database\footnote{http://sagadatabase.jp/}, and the \citet{Lucey2023} CEMP candidates, along with the reported spectroscopic abundances, if available, and references for each star. In total, reference values are listed for 3,923 stars. Of these, 1,082 candidate CEMP stars, cross-matched with \citet{Lucey2023} using Gaia ID, are indicated in the last column with a True or False flag, along with the CEMP-candidate probabilities reported in their study.

Representative parameters and values for stars cross-matched within a 5 arcsecond radius are provided for informative comparison between photometric estimates of [Fe/H] and [C/Fe] and spectroscopic measurements. For example, a subset of the values for the 561 stars in common with L25 (CEMP-no: 108, CEMP-$s$: 453, Group I/II/III/IV: 425, 105, 4, 27) are provided in Table~\ref{tab:jsplus_l25_cemp_stub}. In addition, the 16 CEMP stars in common with JINAbase (CEMP-no: 7, CEMP-$s$: 9, Group I/II/III/IV: 7, 4, 3, 2) are listed in Table~\ref{tab:jsplus_jina_cemp}, and the 36 stars in common with SAGA (CEMP-no: 16, CEMP-$s$: 20, Group I/II/III/IV: 17, 13, 4, 2) are shown in Table~\ref{tab:jsplus_saga_cemp}.

In the tables, the ''Type" column indicates whether a star is a dwarf or a giant, and it also includes data-quality labels based on the coefficient of determination $R^2$ described below, which can facilitate future spectroscopic follow-up. For each stellar type, within each temperature range and for each abundance flag cut (see Table~\ref{tab:phot_spec_r2}), $R^2$ was calculated and used to assign data quality labels as follows: (-) for $R^2 < 0.3$, low (L) for $0.3 \leq R^2 < 0.4$, medium (M) for $0.4 \leq R^2 < 0.5$, and high (H) for $R^2 \geq 0.5$. Then, these labels were assumed to apply to the full combined J/S-PLUS sample. For example, giants were consistently assigned labels (H, H, H) and (H, H, M) for [Fe/H] and [C/Fe]$_{\rm c}$, corresponding to flags $\ge$ 0.85, 0.90, and 0.95, regardless of temperature. For dwarfs, stars with 4,000 K $\leq T_{\rm eff} \leq$ 5,750 K were assigned (-, -, H) for both abundances with flag cuts, stars with 5,750 K $< T_{\rm eff} \leq$ 6,000 K were assigned (-, L, H) and (-, -, M) for [Fe/H] and [C/Fe]$_{\rm c}$, respectively, and the same procedure was applied for the higher temperature cuts at 6,250 K and 6,700 K.

\begin{figure*}
    \centering
    \includegraphics[width=0.8\linewidth]{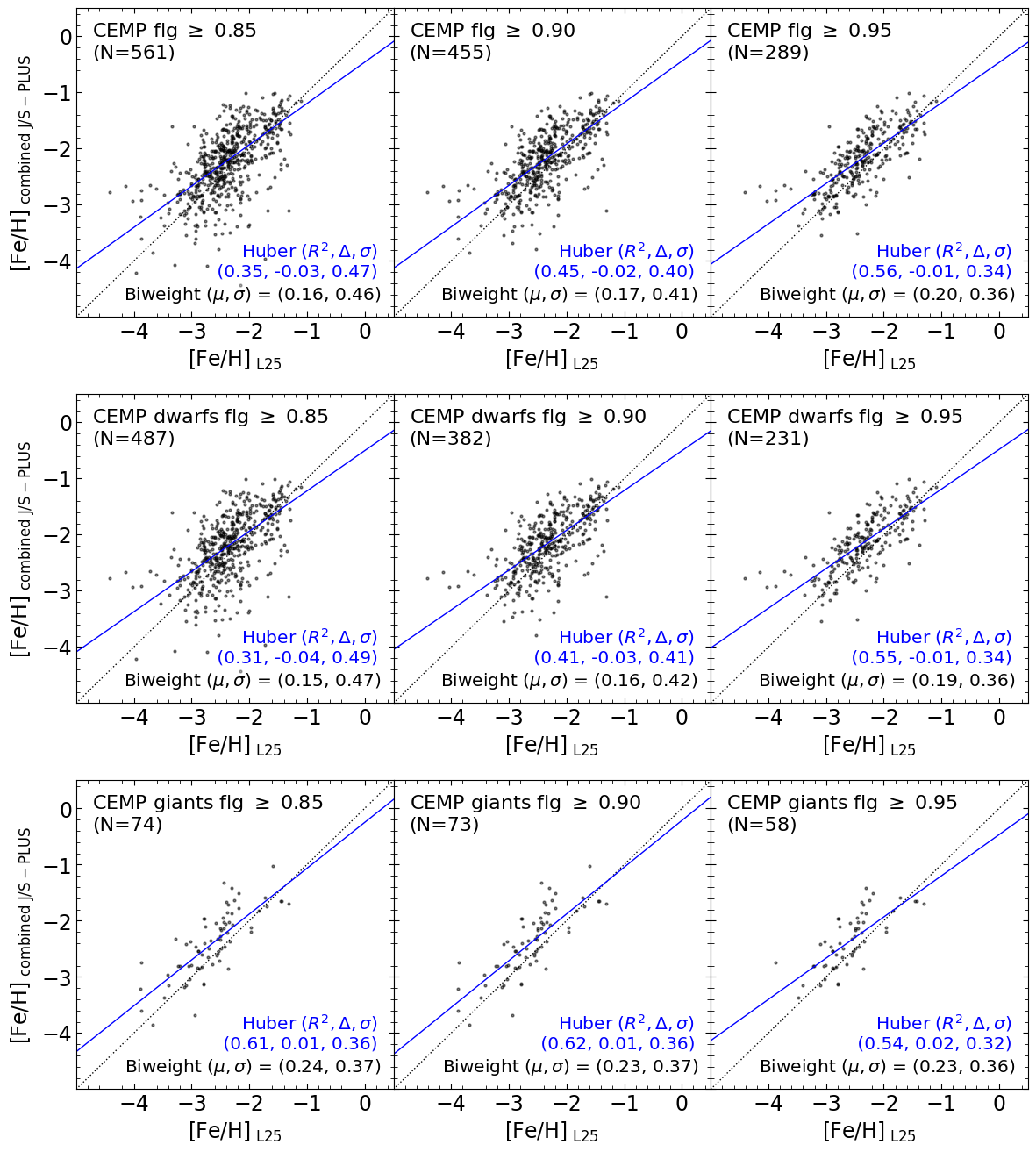}
    \caption{Comparison of photometric [Fe/H] estimates with spectroscopic measurements from L25 for CEMP stars identified in both surveys. Each row shows all stars, dwarfs, and giants. The left, middle, and right panels display the combined J/S-PLUS sample with [Fe/H] and [C/Fe]$_{\rm c}$ flag values of $\geq$ 0.85, 0.90, and 0.95, respectively. The number of stars in each sample is indicated in the top-left corner. Dashed-black lines show the one-to-one relation, while the solid-blue lines represent the Huber linear regression, with the resulting $R^2$, $\Delta$, and $\sigma$ shown in blue at the lower right. The biweight location $\mu$ and scale $\sigma$ of the residuals are also calculated and indicated in black below.}
    \label{Fig:photspecfeh}
\end{figure*}

Figure~\ref{Fig:photspecfeh} shows a comparison between the photometric [Fe/H] estimates from this study and the spectroscopic measurements for stars in common with L25. Each row of panels shows the comparison for all CEMP stars, dwarfs, and giants, with the number of stars in each sample indicated in the top-left corner. The left, middle, and right panels show the combined J/S-PLUS sample, with [Fe/H] and [C/Fe]$_{\rm c}$ flag values of $\geq$ 0.85, 0.90, and 0.95, respectively. 

\begin{figure*}
    \centering
    \includegraphics[width=0.8\linewidth]{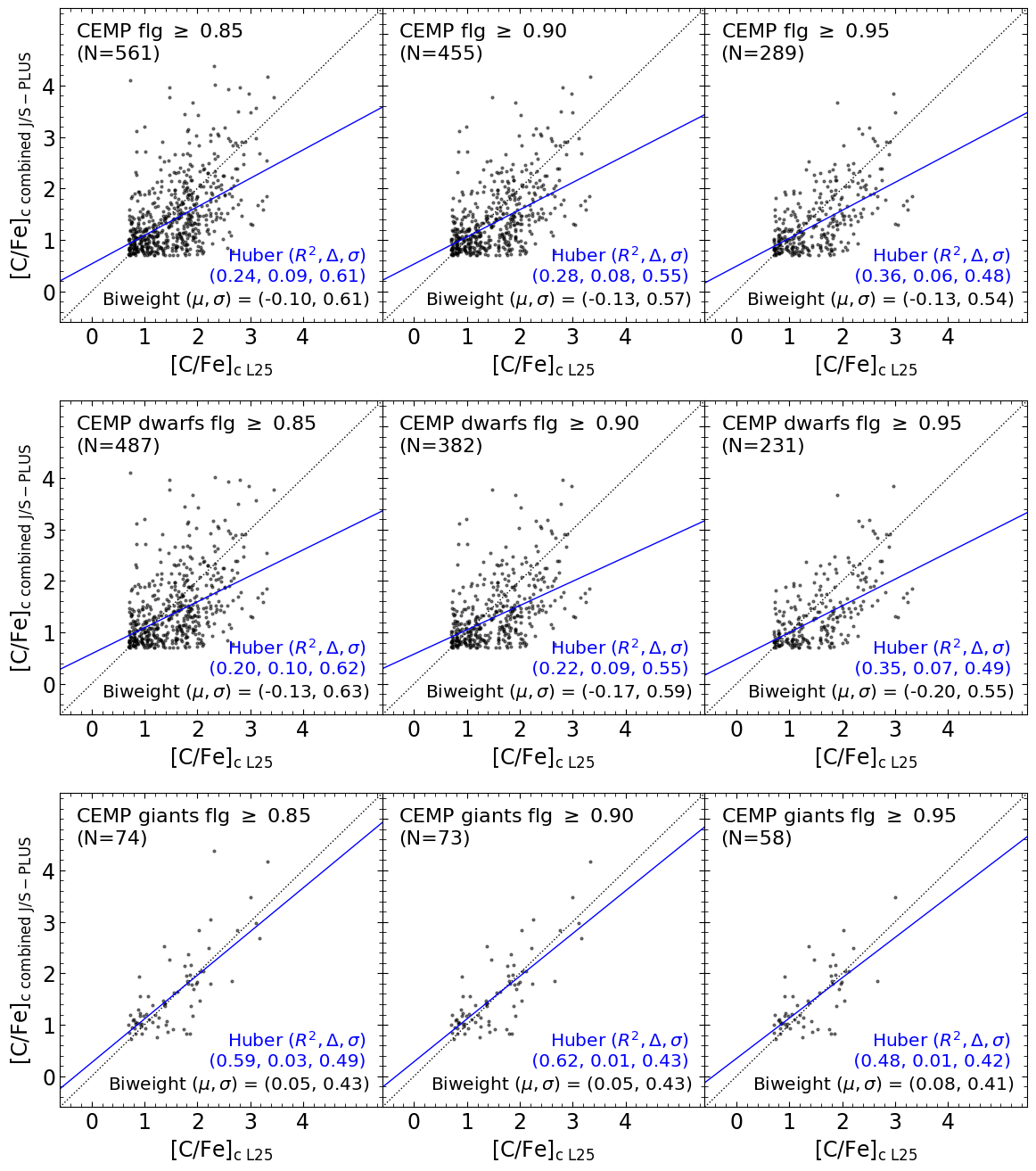}
    \caption{The same as Figure~\ref{Fig:photspecfeh}, but for [C/Fe]$_{\rm c}$.}
    \label{Fig:photspeccfe}
\end{figure*}

In each panel, the dashed-black line denotes the one-to-one relationship, while the solid-blue line shows a Huber regression \citep{Huber1964}, a linear regression model that is robust to outliers. For data points with residuals (observed $−$ predicted) less than or equal to a threshold $\epsilon$, the Huber loss function is proportional to the square of the residual, and the model is optimized to minimize it as in ordinary least squares. For residuals greater than $\epsilon$, the loss increases proportionally to the absolute value of the residual, limiting the influence of outliers and allowing the model to produce stable predictions. In this study, we used the default value $\epsilon = 1.35$ for \texttt{HuberRegressor} in scikit-learn’s \texttt{linear\_model} module. The resulting values of $R^2$, offset $\Delta$, and dispersion $\sigma$ are shown in blue at the lower-right of each panel, and the biweight location $\mu$ and the scale $\sigma$ of the residuals \citep{Beers1990} are indicated in black below.  Figure~\ref{Fig:photspeccfe} is similar to Figure~\ref{Fig:photspecfeh}, but presents a comparison for [C/Fe]$_{\rm c}$.

In general, it is evident that the values of $R^2$ increase as the abundance flag threshold increases. When examining dwarfs and giants separately, the $R^2$ values for dwarfs are comparable to those obtained without distinguishing stellar types, whereas giants generally exhibit $R^2$ values greater than 0.50. For [C/Fe]$_{\rm c}$, the $R^2$ values are approximately 0.1 lower than those for [Fe/H], yet giants still exhibit higher values, indicating a relatively better agreement between photometric and spectroscopic estimates. This is because metallic lines weaken in warmer dwarfs \citep[][Huang et al., submitted]{Yoon2020}, while cooler giants with lower surface gravity exhibit stronger CH $G$-band absorption features. The $R^2$ values shown in the figures, along with the corresponding quality labels L, M, and H, as well as the $R^2$ values and quality labels for dwarf subsamples restricted to $T_{\rm eff} \leq$ 6,250 K, 6,000 K, and 5,750 K, are provided in Table~\ref{tab:phot_spec_r2}.

Table \ref{tab:feh_cfe_mean_err} lists, for each metallicity bin used in Figure \ref{Fig:cempfrac}, the number of stars in the photometric and spectroscopic samples from the combined J/S-PLUS and L25, and also gives the mean values and mean errors of [Fe/H] and [C/Fe]$_{\rm c}$. Values corresponding to the [Fe/H] and [C/Fe] flag thresholds of 0.85, 0.90, and 0.95 are also included.

%\clearpage

\startlongtable
\begin{deluxetable*}{l p{0.65\textwidth} l}
\renewcommand{\arraystretch}{1.05}
%\centering
%\tablewidth{0pt}
\tabletypesize{\scriptsize}
\tablewidth{0.95\textwidth}
%\tablecaption{\label{placeholder}}
\tablecaption{Description of 104,941 Candidate CEMP Stars in the Combined J/S-PLUS Sample \label{tab:colsdescription}}
\tablehead{\colhead{Field} \hspace{1.5cm} & \colhead{Description} \hspace{11.1cm} & \colhead{Unit} \hspace{0.5cm} } 
\startdata
Gaia ID & Gaia DR3 Source ID [gaiaid] & $...$\\
R.A. & R. A. from J-PLUS DR3 and S-PLUS DR4 (J2000) & hours : minutes : seconds\\
Decl. & Decl. from J-PLUS DR3 and S-PLUS DR4 (J2000) & degrees : minutes : seconds\\
$G$mag & Calibration-corrected $G$ magnitude by \citet[][Huang et al. 2025, submitted]{Huang2024} for the Gaia DR3 [gmag] & mag\\
err$_{G~\rm {mag}}$ & Calibration-corrected $G$ magnitude uncertainty by \citet[][Huang et al. 2025, submitted]{Huang2024} for the Gaia DR3 [err\_gmag] & $...$\\
$g$ & $g$ magnitude of combined J/S-PLUS from \citet[][Huang et al. 2025, submitted]{Huang2024} [g] & mag\\
err$_g$ & $g$ magnitude uncertainty of combined J/S-PLUS from \citet[][Huang et al. 2025, submitted]{Huang2024} [err\_g] & $...$\\
$r$ & $r$ magnitude of combined J/S-PLUS from \citet[][Huang et al. 2025, submitted]{Huang2024} [r] & mag\\
err$_r$ & $r$ magnitude uncertainty of combined J/S-PLUS from \citet[][Huang et al. 2025, submitted]{Huang2024} [err\_r] & $...$\\
$g - r$ & $g - r$ color of combined J/S-PLUS [g\_r] & $...$\\
Dist & The distance from \citet[][Huang et al. 2025, submitted]{Huang2024} [dist] & kpc\\
err$_{\rm Dist}$ & Distance uncertainty from \citet[][Huang et al. 2025, submitted]{Huang2024} [err\_dist] & kpc\\
flg$_{\rm Dist}$ & Distance flag from \citet[][Huang et al. 2025, submitted]{Huang2024} [flg\_dist] & $...$\\
 & ``GaiaEDR3" is derived from Gaia parallaxes, while ``CMD\_GIANT", ``CMD\_DWARF\_NOBIAS", and\\ 
 & ``CMD\_DWARF" are obtained from color–absolute-magnitude fiducial relations & $...$\\
E(B-V) & E(B $−$ V) from the extinction map of \citet{Schlegel1998}, corrected by \citet[][Huang et al. 2025, submitted]{Huang2024} [ebv] & $...$\\
$T_{\rm eff}$ & Effective temperature from \citet[][Huang et al. 2025, submitted]{Huang2024} [teff] & K\\
err$_{T_{\rm eff}}$ & Effective temperature uncertainty from \citet[][Huang et al. 2025, submitted]{Huang2024} [err\_teff] & K\\
log~$g$ & Surface gravity from \citet[][Huang et al. 2025, submitted]{Huang2024} [logg] & $...$\\
err$_{{\rm log}~g}$ & Surface gravity uncertainty from \citet[][Huang et al. 2025, submitted]{Huang2024} [err\_logg] & dex\\
Type$_{\rm Huang}$ & Flag as ``D" and ``G" from \citet[][Huang et al. 2025, submitted]{Huang2024} [type\_huang] & $...$\\
Type & Flagged as ``D" and ``G" divided using log $g$ = 3.5, following \citet{Lee2025}, with data quality labels ``–", ``L", ``M", & \\
 & and ``H" obtained from comparison with spectroscopic abundances, each given with [Fe/H] and [C/Fe]$_{\rm c}$ [type] & $...$\\
Subtype$_{\rm Huang}$ & Flag as ``CMD\_MS" for main-sequence stars and ``CMD\_TO" for turn-off stars from \citet[][Huang et al. 2025, submitted]{Huang2024} [subtype\_Huang] & $...$\\
$\rm [Fe/H]$ & Photometric $\rm [Fe/H]$ estimates from \citet[][Huang et al. 2025, submitted]{Huang2024} [feh] & $...$\\
err$_{\rm [Fe/H]}$ & Photometric $\rm [Fe/H]$ estimates uncertainty from \citet[][Huang et al. 2025, submitted]{Huang2024} [err\_feh] & dex\\
flg$_{\rm [Fe/H]}$ & Quality flag of photometric $\rm [Fe/H]$ estimates from \citet[][Huang et al. 2025, submitted]{Huang2024} [flg\_feh] & $...$\\
${\rm [C/Fe]}$ & Photometric $\rm [C/Fe]$ estimates from \citet[][Huang et al. 2025, submitted]{Huang2024} [cfe] & $...$\\
err$_{\rm [C/Fe]}$ & Photometric $\rm [C/Fe]$ estimates uncertainty from \citet[][Huang et al. 2025, submitted]{Huang2024} [err\_cfe] & dex\\
flg$_{\rm [C/Fe]}$ & Quality flag of photometric $\rm [C/Fe]$ estimates from \citet[][Huang et al. 2025, submitted]{Huang2024} [flg\_cfe] & $...$\\
${\rm [C/Fe]_c}$ & Photometric $\rm [C/Fe]$ estimates corrected following \citet{Placco2014c} [cfe\_c] & $...$\\
$A$(C) & Absolute carbon abundance derived from $\rm [C/Fe]$ + $\rm [Fe/H]$ + $\log \epsilon(\mathrm{C})_\odot$, adopting the solar carbon value of 8.43 from \citet{Asplund2009} [ac] & $...$\\
$A$(C)$_{\rm c}$ & Absolute carbon abundance corrected following \citet{Placco2014c} [ac\_c] & $...$\\
Group & Flag as ``I", ``II", ``III", or ``IV" to indicate the assigned CEMP group & $...$\\
Name$_{\rm L25}$ & Star name from \citet{Lee2025} catalog [name\_L25] & $...$\\
Survey$_{\rm L25}$ & Survey name from \citet{Lee2025} catalog, including SDSS, SEGUE1, SEGUE2, BOSS, eBOSS, and LAMOST [survey\_L25] & $...$\\
${\rm [Fe/H]_{L25}}$ & $\rm [Fe/H]$ from \citet{Lee2025} catalog [feh\_L25] & $...$\\
err$_{\rm [Fe/H]_{L25}}$ & $\rm [Fe/H]$ error from \citet{Lee2025} catalog [err\_feh\_L25] & dex\\
${\rm [C/Fe]_{L25}}$ & $\rm [C/Fe]$ from \citet{Lee2025} catalog [cfe\_L25] & $...$\\
err$_{\rm [C/Fe]_{L25}}$ & $\rm [C/Fe]$ error from \citet{Lee2025} catalog [err\_cfe\_L25] & dex\\
${\rm [C/Fe]_{c~L25}}$ & $\rm [C/Fe]_c$ from \citet{Lee2025} catalog [cfe\_c\_L25] & $...$\\
$A$(C)$_{\rm L25}$ & $A$(C) from \citet{Lee2025} catalog [ac\_L25] & $...$\\
$A$(C)$_{\rm c~L25}$ & A(C)$_{\rm c}$ corrected following \citet{Placco2014c} from \citet{Lee2025} catalog [ac\_c\_L25] & $...$\\
Name$_{\rm JINA}$ & Star name from JINAbase [name\_jina] & $...$\\
Ref$_{\rm JINA}$ & Reference(s) for the star from JINAbase [ref\_jina] & $...$\\
${\rm [Fe/H]_{JINA}}$ & $\rm [Fe/H]$ from JINAbase [feh\_jina] & $...$\\
${\rm [C/Fe]_{limit~JINA}}$ & Upper or lower limit of $\rm [C/Fe]$ from JINAbase, indicated by ''$<$" or ''$>$" [cfe\_limit\_jina] & $...$\\
${\rm [C/Fe]_{JINA}}$ & $\rm [C/Fe]$ from JINAbase [cfe\_jina] & $...$\\
$A$(C)$_{\rm JINA}$ & $A$(C) from JINAbase [ac\_jina] & $...$\\
Name$_{\rm SAGA}$ & Star name from the SAGA database [name\_saga] & $...$\\
Ref$_{\rm SAGA}$ & Reference(s) for the star from the SAGA database [ref\_saga] & $...$\\
${\rm [Fe/H]_{SAGA}}$ & $\rm [Fe/H]$ from the SAGA database [feh\_saga] & $...$\\
err$_{\rm [Fe/H]_{SAGA}}$ & $\rm [Fe/H]$ error from the SAGA database [err\_feh\_saga] & dex\\
${\rm [C/Fe]_{SAGA}}$ & $\rm [C/Fe]$ from the SAGA database [cfe\_saga] & $...$\\
err$_{\rm [C/Fe]_{SAGA}}$ & $\rm [C/Fe]$ error from the SAGA database [err\_cfe\_saga] & dex\\
$A$(C)$_{\rm SAGA}$ & $A$(C) from the SAGA database [ac\_saga] & $...$\\
Lucey23 & Flag as "True'' or "False'' to indicate CEMP identification in \citet{Lucey2023} [Lucey23] & $...$\\
Lucey23$_{\rm prob}$ & CEMP probability reported in \citet{Lucey2023} [Lucey23\_prob] & $...$\\
\\[-2.6ex]
\enddata
%\tablecomments{}
\end{deluxetable*}

\startlongtable
\begin{deluxetable*}{ccccccccccccccc}
\tablewidth{0.8\textwidth}
%\tablecaption{\label{placeholder}}
\tablecaption{561 CEMP Candidates in the Combined J/S-PLUS Sample Cross-Matched with L25 \label{tab:jsplus_l25_cemp_stub}}
\tablehead{\colhead{Gaia ID} & \colhead{NAME$_{\rm L25}$} & \colhead{R.A.} & \colhead{Decl.} & \colhead{\textit{G}} & \colhead{Type} & \colhead{[Fe/H]} & \colhead{${\rm [C/Fe]_c}$} & \colhead{[Fe/H]$_{\rm L25}$} & \colhead{[C/Fe]$_{\rm c~L25}$} & \colhead{Group} 
\\[-2ex] 
\colhead{} & \colhead{} & \colhead{(J2000)} & \colhead{(J2000)} & \colhead{(mag)} & \colhead{(F/C)} & \colhead{} & \colhead{} & \colhead{} & \colhead{} & \colhead{}
    }
\startdata
2745479297606453760 & 8740-57367-0705 & 00:00:10.30 & +06:48:31.91 & 17.73 & D (L/-) & $-$3.33 & $+$1.81 & $-$2.43 & $+$1.25 & III \\
2752916295473502464 & EG~000015N080026M01-57309-09053 & 00:03:57.86 & +08:59:20.96 & 15.68 & D (M/-) & $-$2.89 & $+$1.18 & $-$2.91 & $+$2.19 & II \\
2545744527861371392 & 7850-56956-0907 & 00:04:30.63 & -00:35:56.93 & 16.53 & D (H/M) & $-$2.17 & $+$1.90 & $-$2.30 & $+$1.58 & I \\
2876757206391883008 & 7750-58402-0015 & 00:05:04.02 & +34:58:13.51 & 17.30 & D (-/-) & $-$2.55 & $+$1.34 & $-$2.44 & $+$1.61 & I \\
2877332903808934144 & 7747-58398-0701 & 00:08:58.50 & +36:29:31.10 & 17.60 & D (H/L) & $-$2.51 & $+$1.89 & $-$1.56 & $+$1.08 & I \\
2751799638336865792 & EG~001605N080655B01-56919-10231 & 00:10:04.74 & +07:56:06.08 & 15.66 & G (H/H) & $-$3.68 & $+$3.04 & $-$2.94 & $+$2.25 & IV \\
2876082415490652672 & 7126-56568-0836 & 00:10:57.31 & +35:11:38.52 & 17.29 & D (M/L) & $-$1.83 & $+$1.15 & $-$2.66 & $+$1.78 & I \\
2876080323842041216 & M~31002N36B1-57011-08206 & 00:11:28.60 & +35:10:49.37 & 15.67 & D (H/M) & $-$2.05 & $+$0.90 & $-$2.29 & $+$1.59 & I \\
2741125785612463616 & 8745-57391-0795 & 00:17:22.03 & +04:33:57.32 & 16.42 & D (H/M) & $-$2.41 & $+$1.50 & $-$2.48 & $+$1.69 & I \\
2546861769114476032 & 9406-58067-0015 & 00:20:31.38 & +00:27:11.27 & 17.31 & D (H/M) & $-$2.35 & $+$1.72 & $-$2.48 & $+$1.60 & I \\
2547133176688396800 & 4300-55528-0244 & 00:20:50.00 & +01:17:44.73 & 17.23 & D (L/-) & $-$3.09 & $+$2.03 & $-$2.29 & $+$1.81 & I \\
366303780558169344 & GACII~004N36B1-58040-08130 & 00:23:40.04 & +35:58:04.13 & 13.98 & G (H/M) & $-$2.77 & $+$1.38 & $-$2.64 & $+$1.15 & II \\
2543588316840465408 & 3586-55181-0278 & 00:31:59.53 & -00:11:13.20 & 17.04 & D (H/M) & $-$2.72 & $+$2.02 & $-$2.64 & $+$2.34 & I \\
2542843672591011840 & 9404-58045-0862 & 00:33:51.33 & -00:04:08.23 & 17.12 & D (H/M) & $-$1.54 & $+$1.37 & $-$1.55 & $+$0.74 & I \\
2544000015225846784 & 1134-52644-442 & 00:34:20.20 & +00:25:51.96 & 16.90 & D (H/M) & $-$2.07 & $+$1.89 & $-$1.60 & $+$0.92 & I \\
2555687755108422016 & 2312-53709-201 & 00:35:09.08 & +06:20:20.28 & 15.34 & D (H/H) & $-$1.82 & $+$1.64 & $-$2.09 & $+$1.20 & I \\
2544131402568610304 & 7868-57006-0756 & 00:36:16.68 & +00:50:05.01 & 17.01 & D (H/M) & $-$2.11 & $+$0.83 & $-$2.13 & $+$0.86 & I \\
2809148545637525632 & EG~004228N273834M01-58070-05212 & 00:39:28.26 & +26:24:14.32 & 17.06 & D (L/-) & $-$2.43 & $+$1.12 & $-$2.31 & $+$1.41 & II \\
2549397307351168512 & 1495-52944-368 & 00:40:30.18 & +01:07:23.96 & 17.34 & D (M/-) & $-$2.41 & $+$0.97 & $-$2.52 & $+$1.17 & II \\
2551078701148580864 & 4303-55508-0028 & 00:40:56.11 & +03:20:21.36 & 16.82 & D (M/L) & $-$2.00 & $+$1.04 & $-$2.49 & $+$1.92 & I \\
\\[-2.6ex]
\enddata
\tablecomments{This table is a stub; the full table is available in the electronic edition.}
\end{deluxetable*}

\startlongtable
\begin{deluxetable*}{ccccccccccccccc}
\tablewidth{0.8\textwidth}
%\tablecaption{\label{placeholder}}
\tablecaption{16 CEMP Candidates in the Combined J/S-PLUS Sample Cross-Matched with JINAbase \label{tab:jsplus_jina_cemp}}
\tablehead{\colhead{Gaia ID} & \colhead{NAME$_{\rm JINA}$} & \colhead{R.A.} & \colhead{Decl.} & \colhead{\textit{G}} & \colhead{Type} & \colhead{[Fe/H]} & \colhead{${\rm [C/Fe]_c}$} & \colhead{[Fe/H]$_{\rm JINA}$} & \colhead{[C/Fe]$_{\rm c~JINA}$} & \colhead{Group} 
\\[-2ex] 
\colhead{} & \colhead{} & \colhead{(J2000)} & \colhead{(J2000)} & \colhead{(mag)} & \colhead{(F/C)} & \colhead{} & \colhead{} & \colhead{} & \colhead{} & \colhead{}
    }
\startdata
4952088844089190400 & HE~0231-4016 & 02:33:44.42 & -40:03:42.58 & 15.94 & D (H/M) & $-$2.14 & $+$1.59 & $-$2.08 & $+$1.32 & I \\
5065476358659953664 & HE~0243-3044 & 02:45:16.44 & -30:32:02.14 & 15.89 & G (H/H) & $-$2.63 & $+$5.06 & $-$2.58 & $+$2.43 & I \\
5052420242197525504 & HE~0251-3216 & 02:53:05.52 & -32:04:22.14 & 15.23 & D (L/-) & $-$1.75 & $+$1.55 & $-$3.15 & $+$2.44 & I \\
4856059422664301568 & HE~0338-3945 & 03:39:55.02 & -39:35:43.08 & 15.24 & D (H/M) & $-$2.32 & $+$1.48 & $-$2.42 & $+$2.06 & I \\
4786891482623486976 & HE~0450-4705 & 04:51:33.60 & -47:00:04.50 & 14.15 & G (H/M) & $-$3.13 & $+$0.74 & $-$3.10 & $+$0.88 & II \\
4786258163926019072 & HE~0450-4902 & 04:51:43.21 & -48:57:25.39 & 15.76 & D (H/L) & $-$2.35 & $+$1.28 & $-$3.07 & $+$2.03 & I \\
897312989912922496 & SDSS~J072352.21+363757.2 & 07:23:52.21 & +36:37:57.19 & 14.90 & G (H/M) & $-$3.50 & $+$2.05 & $-$3.32 & $+$1.79 & III \\
812808546291531520 & SDSS~0924+40 & 09:24:01.81 & +40:59:28.00 & 15.42 & D (H/M) & $-$2.97 & $+$2.98 & $-$2.56 & $+$2.73 & I \\
5674504400563869696 & HE~1005-1439 & 10:07:52.42 & -14:54:21.43 & 13.57 & G (H/H) & $-$3.19 & $+$2.22 & $-$3.22 & $+$2.49 & IV \\
3976087728282022272 & SDSS~J114323.42+202058.0 & 11:43:23.40 & +20:20:57.92 & 16.56 & D (M/-) & $-$2.59 & $+$1.14 & $-$3.13 & $+$2.80 & II \\
3522792310017214464 & HE~1245-1616 & 12:47:56.77 & -16:32:44.17 & 16.11 & D (M/-) & $-$2.98 & $+$1.57 & $-$2.98 & $+$0.81 & II \\
3658222348370838528 & SDSS~J134913.54-022942.8 & 13:49:13.53 & -02:29:42.84 & 16.30 & D (H/M) & $-$3.24 & $+$2.55 & $-$3.14 & $+$2.96 & IV \\
3632963817501130240 & HE~1346-0427 & 13:49:24.94 & -04:42:14.38 & 13.99 & D (L/-) & $-$3.45 & $+$1.14 & $-$3.57 & $+$1.10 & II \\
1428076733295455232 & SDSS~J161313.53+530909.7 & 16:13:13.51 & +53:09:09.72 & 16.22 & G (H/M) & $-$3.15 & $+$1.81 & $-$3.32 & $+$2.09 & III \\
1348458828585714816 & SDSS~J173417.89+431606.5 & 17:34:17.89 & +43:16:06.41 & 16.01 & G (H/M) & $-$2.83 & $+$2.43 & $-$2.51 & $+$1.78 & I \\
6788448668941293568 & CS~29498-043 & 21:03:52.10 & -29:42:50.30 & 13.35 & G (H/H) & $-$3.76 & $+$1.93 & $-$3.81 & $+$2.38 & III \\
\\[-2.6ex]
\enddata
%\tablecomments{This table is a stub; the full table is available in the electronic edition.}
\end{deluxetable*}
\startlongtable
\begin{deluxetable*}{ccccccccccccccc}
\tablewidth{0.8\textwidth}
%\tablecaption{\label{placeholder}}
\tablecaption{36 CEMP Candidates in the Combined J/S-PLUS Sample Cross-Matched with SAGA \label{tab:jsplus_saga_cemp}}
\tablehead{\colhead{Gaia ID} & \colhead{NAME$_{\rm SAGA}$} & \colhead{R.A.} & \colhead{Decl.} & \colhead{\textit{G}} & \colhead{Type} & \colhead{[Fe/H]} & \colhead{${\rm [C/Fe]_c}$} & \colhead{[Fe/H]$_{\rm SAGA}$} & \colhead{[C/Fe]$_{\rm c~SAGA}$} & \colhead{Group} 
\\[-2ex] 
\colhead{} & \colhead{} & \colhead{(J2000)} & \colhead{(J2000)} & \colhead{(mag)} & \colhead{(F/C)} & \colhead{} & \colhead{} & \colhead{} & \colhead{} & \colhead{}
    }
\startdata
2546455980604158464 & 2MASS~J00114258+0109386 & 00:11:42.58 & +01:09:38.81 & 14.06 & G (H/M) & $-$2.58 & $+$1.70 & $-$2.18 & $+$0.91 & I \\
2547126442179677952 & HE~0017+0055 & 00:20:21.61 & +01:12:06.64 & 11.25 & G (H/M) & $-$2.49 & $+$1.91 & $-$2.45 & $+$2.44 & I \\
366303780558169344 & J0023+3558 & 00:23:40.04 & +35:58:04.13 & 13.98 & G (H/M) & $-$2.77 & $+$1.38 & $-$2.56 & $+$0.80 & II \\
2533491432841892864 & J0119-0121 & 01:19:39.31 & -01:21:49.70 & 14.99 & D (H/L) & $-$2.85 & $+$2.25 & $-$3.19 & $+$2.42 & I \\
4952088844089190400 & HE~0231-4016 & 02:33:44.42 & -40:03:42.58 & 15.94 & D (H/M) & $-$2.14 & $+$1.59 & $-$2.08 & $+$1.31 & I \\
5065476358659953664 & HE~0243-3044 & 02:45:16.44 & -30:32:02.14 & 15.89 & G (H/H) & $-$2.63 & $+$5.06 & $-$2.58 & $+$2.43 & I \\
2498120964114484736 & J0251-0006 & 02:51:50.30 & -00:06:37.70 & 13.02 & D (H/M) & $-$2.82 & $+$1.34 & $-$2.97 & $+$1.51 & II \\
5052420242197525504 & HE~0251-3216 & 02:53:05.52 & -32:04:22.14 & 15.23 & D (L/-) & $-$1.75 & $+$1.55 & $-$3.29 & $+$2.53 & I \\
4856059422664301568 & HE~0338-3945 & 03:39:55.02 & -39:35:43.08 & 15.24 & D (H/M) & $-$2.32 & $+$1.48 & $-$2.42 & $+$2.11 & I \\
4786891482623486976 & HE~0450-4705 & 04:51:33.60 & -47:00:04.50 & 14.15 & G (H/M) & $-$3.13 & $+$0.74 & $-$3.04 & $+$0.88 & II \\
4786258163926019072 & HE~0450-4902 & 04:51:43.21 & -48:57:25.39 & 15.76 & D (H/L) & $-$2.35 & $+$1.28 & $-$3.07 & $+$2.03 & I \\
946923298155521024 & J0703+3747 & 07:03:01.50 & +37:47:21.79 & 12.98 & D (M/-) & $-$2.54 & $+$0.74 & $-$2.42 & $+$0.98 & II \\
883042050539140992 & LAMOST~J070542.30+255226.6 & 07:05:42.30 & +25:52:26.52 & 13.32 & G (H/H) & $-$3.18 & $+$1.65 & $-$3.26 & $+$1.61 & II \\
899693810543540096 & J0723+3806 & 07:23:23.86 & +38:06:13.33 & 14.47 & D (H/L) & $-$2.81 & $+$0.72 & $-$3.05 & $+$1.00 & II \\
897312989912922496 & SDSS~J0723+3637 & 07:23:52.21 & +36:37:57.19 & 14.90 & G (H/M) & $-$3.50 & $+$2.05 & $-$3.32 & $+$1.79 & III \\
881641444523743488 & J0748+3209 & 07:48:21.55 & +32:09:01.06 & 12.79 & G (H/M) & $-$3.13 & $+$1.85 & $-$3.08 & $+$0.98 & III \\
902428295962842496 & J0814+3305 & 08:14:13.14 & +33:05:57.38 & 14.88 & G (H/M) & $-$3.31 & $+$1.06 & $-$3.34 & $+$0.99 & II \\
708825223391675648 & J0824+3025 & 08:24:59.33 & +30:25:41.75 & 15.93 & D (H/L) & $-$2.71 & $+$1.77 & $-$2.71 & $+$2.21 & I \\
812808546291531520 & SDSS~0924+40 & 09:24:01.81 & +40:59:28.00 & 15.42 & D (H/M) & $-$2.97 & $+$2.98 & $-$2.56 & $+$2.73 & I \\
3828359167040220160 & J1003-0358 & 10:03:57.86 & -03:58:52.19 & 13.46 & D (H/L) & $-$2.12 & $+$0.71 & $-$2.18 & $+$1.31 & II \\
5674504400563869696 & HE~1005-1439 & 10:07:52.42 & -14:54:21.43 & 13.57 & G (H/H) & $-$3.19 & $+$2.22 & $-$3.22 & $+$2.49 & IV \\
804170782944450048 & J1020+4046 & 10:20:59.76 & +40:46:53.89 & 14.80 & D (H/M) & $-$2.80 & $+$1.91 & $-$2.93 & $+$2.37 & I \\
3749925058396764160 & HE~1045-1434 & 10:47:44.20 & -14:50:22.52 & 14.52 & G (H/H) & $-$2.61 & $+$2.41 & $-$2.55 & $+$3.21 & I \\
3976087728282022272 & SDSS~J1143+2020 & 11:43:23.40 & +20:20:57.92 & 16.56 & D (M/-) & $-$2.59 & $+$1.14 & $-$3.13 & $+$2.80 & II \\
3522792310017214464 & HE~1245-1616 & 12:47:56.77 & -16:32:44.17 & 16.11 & D (M/-) & $-$2.98 & $+$1.57 & $-$2.92 & $+$0.91 & II \\
3658222348370838528 & SDSS~J1349-0229 & 13:49:13.53 & -02:29:42.84 & 16.30 & D (H/M) & $-$3.24 & $+$2.55 & $-$3.07 & $+$3.04 & IV \\
1450878027474744064 & J1401+2659 & 14:01:18.72 & +26:59:50.71 & 13.25 & D (H/L) & $-$2.86 & $+$0.90 & $-$3.05 & $+$1.10 & II \\
1610438398983648256 & SDSS~J142441.88+561535.0 & 14:24:41.83 & +56:15:34.99 & 15.56 & D (H/L) & $-$2.76 & $+$1.07 & $-$3.06 & $+$1.26 & II \\
1390760133282619136 & 2MASS~J15312547+4220551 & 15:31:25.47 & +42:20:55.10 & 15.01 & G (H/H) & $-$2.77 & $+$1.55 & $-$2.08 & $+$0.83 & I \\
1204509907284481792 & J1551+2040 & 15:51:50.41 & +20:40:33.17 & 14.37 & G (H/M) & $-$2.84 & $+$2.27 & $-$2.48 & $+$0.75 & I \\
1428076733295455232 & SDSS~J1613+5309 & 16:13:13.51 & +53:09:09.72 & 16.22 & G (H/M) & $-$3.15 & $+$1.81 & $-$3.32 & $+$2.09 & III \\
1348458828585714816 & SDSS~J1734+4316 & 17:34:17.89 & +43:16:06.41 & 16.01 & G (H/M) & $-$2.83 & $+$2.43 & $-$2.51 & $+$1.78 & I \\
4227964882766196224 & SDSS~2047+00 & 20:47:28.87 & +00:15:53.37 & 16.16 & D (H/M) & $-$1.79 & $+$0.95 & $-$2.10 & $+$2.01 & I \\
6788448668941293568 & CS~29498-043 & 21:03:52.10 & -29:42:50.30 & 13.35 & G (H/H) & $-$3.76 & $+$1.93 & $-$3.78 & $+$2.47 & III \\
2653114101760354304 & SDSS~J2239-0048 & 22:39:46.45 & -00:48:27.79 & 15.41 & D (M/L) & $-$1.12 & $+$0.80 & $-$1.47 & $+$0.95 & I \\
2718453065572920576 & J2253+1134 & 22:53:11.52 & +11:34:10.89 & 13.52 & G (H/M) & $-$2.54 & $+$1.17 & $-$2.51 & $+$0.78 & II \\
\\[-2.6ex]
\enddata
%\tablecomments{This table is a stub; the full table is available in the electronic edition.}
\end{deluxetable*}

% \onecolumngrid
% \newpage
% \twocolumngrid

\startlongtable
\begin{deluxetable*}{lccccccc}
\renewcommand{\arraystretch}{1}
\tabletypesize{\footnotesize}
\LTcapwidth=\textwidth
\setlength{\tabcolsep}{6pt}
\tablecaption{$R^2$ Values of Photometric and Spectroscopic CEMP stars \label{tab:phot_spec_r2}}
\tablehead{\colhead{$T_{\rm eff}$} & \colhead{Type} & \colhead{} & \colhead{[Fe/H]} & \colhead{} & \colhead{} & \colhead{[C/Fe]$_{\rm c}$} & \colhead{} 
\\[-2ex] 
\colhead{(K)} & \colhead{} & \colhead{flg $\geq$ 0.85} & \colhead{$\geq$ 0.90} & \colhead{$\geq$ 0.95} & \colhead{flg $\geq$ 0.85} & \colhead{$\geq$ 0.90} & \colhead{$\geq$ 0.95}
    }
\startdata
% -------------------- All --------------------
\\[-2.5ex]
All & All
& ~0.35 (L)\tablenotemark{\scriptsize a} & 0.45 (M) & 0.56 (H) & 0.24 (--) & 0.28 (--) & 0.36 (L) \\
& Dwarfs
& 0.31 (L) & 0.41 (M) & 0.55 (H) & 0.20 (--) & 0.22 (--) & 0.35 (L) \\
& Giants
& 0.61 (H) & 0.62 (H) & 0.54 (H) & 0.59 (H) & 0.62 (H) & 0.48 (M) \\
\\[-2.5ex]
\hline
% -------------------- Teff <= 6250 --------------------
\\[-2.5ex]
$\leq 6,250$ 
& Dwarfs
& 0.31 (L) & 0.40 (M) & 0.56 (H) & 0.23 (--) & 0.24 (L) & 0.42 (M) \\
\\[-2.5ex]
\hline
% -------------------- Teff <= 6000 --------------------
\\[-2.5ex]
$\leq 6,000$ 
& Dwarfs
& 0.28 (--) & 0.35 (L) & 0.65 (H) & 0.23 (--) & 0.26 (--) & 0.48 (M) \\
\\[-2.5ex]
\hline
% -------------------- Teff <= 5750 --------------------
\\[-2.5ex]
$\leq 5,750$ 
& Dwarfs
& 0.23 (--) & 0.23 (--) & 0.78 (H) & 0.24 (--) & 0.27 (--) & 0.63 (H) \\
\\[-2.5ex]
\enddata
%\tablecomments{}
\vspace{0.7em}
$^{a}$ The data quality labels based on the $R^2$ are assigned as follows: (–) for $R^2 < 0.3$, low (L) for $0.3 \le R^2 < 0.4$, medium (M) for $0.4 \le R^2 < 0.5$, and high (H) for $R^2 \ge 0.5$.
\end{deluxetable*}

\startlongtable
\begin{deluxetable*}{ccccccc}
\renewcommand{\arraystretch}{1}
\tabletypesize{\footnotesize}
\LTcapwidth=\textwidth
\setlength{\tabcolsep}{6pt}
\tablecaption{Abundances and Errors of Photometric and Spectroscopic CEMP Stars \label{tab:feh_cfe_mean_err}}
\tablehead{\colhead{} & \colhead{flg $\geq$} & \colhead{[Fe/H] $\leq$} & \colhead{$<{\rm [Fe/H]}>$} & \colhead{$<{\rm err_{[Fe/H]}}>$} & \colhead{$<{\rm [C/Fe]}>$} & \colhead{$<{\rm err_{[C/Fe]}}>$}
\\[-2ex] 
\colhead{} & \colhead{} & \colhead{} & \colhead{} & \colhead{(dex)} & \colhead{} & \colhead{(dex)}
    }
\startdata
\multirow{7}{*}{    }  &      & $-$1.0 (104,941) & $-$1.80 & 0.46 & $+$1.13 & 0.50 \\
                       &      & $-$1.5 (63,898) & $-$2.16 & 0.46 & $+$1.23 & 0.52 \\
                       &      & $-$2.0 (33,318) & $-$2.56 & 0.47 & $+$1.38 & 0.53 \\
                       & 0.85 & $-$2.5 (15,129)  & $-$2.95 & 0.48 & $+$1.58 & 0.55 \\
                       &      & $-$3.0 (5,157)   & $-$3.40 & 0.50 & $+$1.93 & 0.58 \\
                       &      & $-$3.5 (1,495)   & $-$3.89 & 0.52 & $+$2.40 & 0.60 \\
                       &      & $-$4.0 (422)     & $-$4.38 & 0.55 & $+$2.90 & 0.63 \\
\noalign{\vskip 2mm}
\multirow{7}{*}{Combined J/S-PLUS}  &      & $-$1.0 (50,202) & $-$1.78 & 0.39 & $+$1.07 & 0.46 \\
                                    &      & $-$1.5 (30,188) & $-$2.14 & 0.39 & $+$1.16 & 0.47 \\
                                    &      & $-$2.0 (15,747)  & $-$2.52 & 0.40 & $+$1.29 & 0.49 \\
                                    & 0.90 & $-$2.5 (6,909)  & $-$2.88 & 0.40 & $+$1.46 & 0.50 \\
                                    &      & $-$3.0 (1,906)   & $-$3.32 & 0.42 & $+$1.85 & 0.52 \\
                                    &      & $-$3.5 (406)     & $-$3.80 & 0.43 & $+$2.44 & 0.55 \\
                                    &      & $-$4.0 (77)     & $-$4.32 & 0.47 & $+$3.11 & 0.60 \\
\noalign{\vskip 2mm}
\multirow{7}{*}{    }  &      & $-$1.0 (10,121) & $-$1.86 & 0.30 & $+$1.01 & 0.41 \\
                       &      & $-$1.5 (6,694) & $-$2.17 & 0.30 & $+$1.09 & 0.42 \\
                       &      & $-$2.0 (3,743)  & $-$2.50 & 0.31 & $+$1.17 & 0.42 \\
                       & 0.95 & $-$2.5 (1,731)   & $-$2.82 & 0.32 & $+$1.27 & 0.43 \\
                       &      & $-$3.0 (353)     & $-$3.21 & 0.33 & $+$1.65 & 0.44 \\
                       &      & $-$3.5 (41)      & $-$3.66 & 0.31 & $+$2.56 & 0.41 \\
                       &      & $-$4.0 (1)       & $-$5.18 & 0.36 & $+$2.29 & 0.68 \\
\midrule
\multirow{7}{*}{L25 Spectroscopic}   &      & $-$1.0 (300,638) & $-$1.54 & 0.05 & $-$0.05 & 0.13 \\
                       &      & $-$1.5 (140,819) & $-$1.89 & 0.05 & $+$0.00 & 0.15 \\
                       &      & $-$2.0 (41,848)  & $-$2.31 & 0.05 & $+$0.19 & 0.18 \\
                       &      & $-$2.5 (8,651)   & $-$2.77 & 0.07 & $+$0.60 & 0.18 \\
                       &      & $-$3.0 (1,399)   & $-$3.25 & 0.13 & $+$1.07 & 0.17 \\
                       &      & $-$3.5 (199)     & $-$3.75 & 0.23 & $+$1.45 & 0.15 \\
                       &      & $-$4.0 (28)      & $-$4.18 & 0.25 & $+$2.14 & 0.13 \\
\enddata

\end{deluxetable*}

\end{appendix}

\vfill\eject

\gdef\thefigure{\thesection.\arabic{figure}}    

\bibliography{main}{}
\bibliographystyle{aasjournalv7}

\end{document}